\documentclass[11pt]{article}
\pdfoutput=1
\usepackage{geometry}                
\usepackage{graphicx}
\usepackage{amssymb}
\usepackage{amsmath}
\usepackage{epstopdf}
\usepackage{amscd}
\usepackage{mathtools}
\usepackage{slashed}
\usepackage{bbm}
\usepackage{upgreek}
\usepackage[utf8]{inputenc}
\usepackage[matrix,arrow]{xy}
\usepackage{arydshln}
\usepackage{MnSymbol}
\usepackage[utf8]{inputenc}

\usepackage{epsf,latexsym,amsthm,bbm}

\usepackage{stolenstyle}

\usepackage[linktocpage=true]{hyperref}
\hypersetup{
colorlinks=true,
citecolor=blue,
linkcolor=blue,
urlcolor=blue}

\DeclareFontFamily{U}{rsf}{}
\DeclareFontShape{U}{rsf}{m}{n}{
  <5> <6> rsfs5 <7> <8> <9> rsfs7 <10-> rsfs10}{}
\DeclareMathAlphabet\Scr{U}{rsf}{m}{n}

\def\CO#1#2{{[#1,#2]}}

\def\iden{{\mathbbm 1}}

\def\R{{\mathbb R}}

\def\Tr{\operatorname{Tr}}

\def\diag{\operatorname{diag}}

\def\tr{\operatorname{tr}}

\def\SO{\operatorname{SO}}

\def\Sp{\operatorname{Sp}}

\def\p{\partial}

\def\la{\langle}
\def\ra{\rangle}

\def\ff#1#2{{\textstyle\frac{#1}{#2}}}
\def\half{\frac{1}{2}}

\def\cF{{\cal F}}
\def\cG{{\cal G}}

\def\cI{{\cal I}}

\newcommand\pih{\widehat{\pi}}

\newcommand\phih{\widehat{\phi}}

\newcommand\vphi{\varphi}

\newcommand\ah{\widehat{a}}

\newcommand\fh{\widehat{f}}

\newcommand\ph{\widehat{p}}
\newcommand\qh{\widehat{q}}

\newcommand\Hh{\widehat{H}}

\newcommand\Ph{\widehat{P}}

\newcommand\Uh{\widehat{U}}

\newcommand\Xh{\widehat{X}}

\DeclareFontFamily{OT1}{pzc}{}
\DeclareFontShape{OT1}{pzc}{m}{it}{<-> s * [1.10] pzcmi7t}{}
\DeclareMathAlphabet{\mathpzc}{OT1}{pzc}{m}{it}

\def\ie{{\it i.e.}~}
\def\eg{{\it e.g.}~}

\def\etc{{\it etc.}~}

\def\Tr{ \, \textrm{Tr} \, }

\def\pd{\partial}
\def\diag{ \, \textrm{diag}}

\def\gammabar{\overbar{\gamma}}

\def\varpibar{\overbar{\varpi}}
\def\nubar{\overbar{\nu}}
\def\lambdabar{\overbar{\lambda}}
\def\phibar{\overbar{\phi}}
\def\Pbar{\overbar{P}}

\def\varphibar{\overbar{\varphi}}
\def\nubar{\overbar{\nu}}
\def\lambdabar{\overbar{\lambda}}
\def\pibar{\overbar{\pi}}
\def\abar{\overbar{\beta}}

\def\half{\frac{1}{2}}

\def\BB{\mathcal{B}}

\def\OO{\mathcal{O}}

\def\ZZ{\mathcal{Z}}

\def\MM{\mathcal{M}}
\def\PP{\mathcal{P}}

\def\II{\mathcal{I}}

\def\MM{\mathcal{M}}
\def\KK{\mathcal{K}}

\def\FF{\mathcal{F}}

\newcommand{\overbar}[1]{\mkern 1.5mu\overline{\mkern-1.5mu#1\mkern-1.5mu}\mkern 1.5mu}

\def\bq{{\boldsymbol{q}}}
\def\bp{{\boldsymbol{p}}}
\def\bnu{{\boldsymbol{\nu}}}

\def\bz{{\boldsymbol{z}}}
\def\bqh{{\widehat{\bq}}}
\def\bph{{\widehat{\bp}}}

\def\bV{{\boldsymbol{V}}}
\def\bA{{\boldsymbol{A}}}
\def\bB{{\boldsymbol{B}}}
\def\bD{{\boldsymbol{D}}}
\def\buppsi{{\boldsymbol{\uppsi}}}

\def\cbM{{\boldsymbol{\mathcal{M}}}}
\def\cbB{{\boldsymbol{\mathcal{B}}}}
\def\cbK{{\boldsymbol{\mathcal{K}}}}
\def\cbH{{\boldsymbol{\mathcal{H}}}}
\def\cbJ{{\boldsymbol{\mathcal{J}}}}
\def\cbS{{\boldsymbol{\mathcal{S}}}}
\def\cbC{{\boldsymbol{\mathcal{C}}}}
\def\cbN{{\boldsymbol{\mathcal{N}}}}
\def\cbR{{\boldsymbol{\mathcal{R}}}}
\def\cbA{{\boldsymbol{\mathcal{A}}}}

\def\bM{{\boldsymbol{M}}}
\def\bC{{\boldsymbol{C}}}
\def\boeta{{\boldsymbol{\eta}}}
\def\bpi{{\boldsymbol{\pi}}}
\def\bxi{{\boldsymbol{\xi}}}
\def\bDelta{{\boldsymbol{\Delta}}}

\def\bVh{{\widehat{\bV}}}
\def\bah{{\widehat{\boldsymbol{a}}}}

\def\tuppsi{{\widetilde{\uppsi}}}
\def\trho{{\widetilde{\rho}}}

\def\tomega{{\widetilde{\omega}}}

\renewcommand{\ra}{\rrangle}
\renewcommand{\la}{\llangle}

\def\preprint{}

\title{Accelerating Solitons}
\author[a] {Ilarion V.~Melnikov,}
\author[b] {Constantinos Papageorgakis,}
\author[c] {Andrew B.~Royston}

\abstract{We present the saddle-point approximation for the effective Hamiltonian of the quantum kink in two-dimensional linear sigma models to all orders in the time-derivative expansion.  We show how the effective Hamiltonian can be used to obtain semiclassical soliton form factors, valid at momentum transfers of order the soliton mass.  Explicit results, however, hinge on finding an explicit solution to a new wave-like partial differential equation, with a time-dependent velocity and a forcing term that depend on the solution.  In the limit of small momentum transfer, the effective Hamiltonian reduces to the expected form, namely $H = \sqrt{P^2 + M^2}$, where $M$ is the one-loop corrected soliton mass, and soliton form factors are given in terms of Fourier transforms of the corresponding classical profiles.} 

\affiliation[a]{Department of Physics and Astronomy, James Madison University\\ 901 Carrier Drive, Harrisonburg, VA 22807, USA}
\affiliation[b]{CRST and School of Physics and Astronomy, Queen Mary University of London\\ Mile End Road, London E1 4NS, UK}
\affiliation[c]{Department of Physics, Penn State Fayette, The Eberly Campus \\ 2201 University Drive, Lemont Furnace, PA 15456, USA}

\emailAdd{melnikix@jmu.edu}
\emailAdd{c.papageorgakis@qmul.ac.uk}
\emailAdd{abr84@psu.edu}

\begin{document}

\begin{flushright}
\preprint{QMUL-PH-20-17}
\end{flushright}

\maketitle
\parskip 7pt

\section{Introduction and Motivation}

The description of soliton states in quantum field theory---the foundations of which were laid out in the mid '70s---is a beautiful subject where basic notions of quantum field theory operate in the background of exact solutions to nonlinear differential equations; for popular reviews see \cite{Jackiw:1977yn,Faddeev:1977rm,Rajaraman:1982is}.  Two-dimensional models possessing kink solitons hold a privileged position:  one can do more analytically, owing to the relative simplicity of working in one spatial dimension and the absence of gauge fields.  

For example, in a class of linear sigma models, including $\phi^4$ theory and other non-integrable models, the exact canonical transformation of phase-space path integration variables from the perturbative sector to the one-soliton sector of the theory is known.  This is a transformation $(\phi(x);\pi(x)) \mapsto (X,\varphi(\rho);P,\varpi(\rho))$, which extracts the soliton collective coordinate $X$ and its conjugate momentum $P$ as a single degree of freedom in the field theory.  The remaining field-theoretic degrees of freedom are collected in a field $\varphi$ and its conjugate $\varpi$ containing fluctuations around the soliton, in such a way that the transformation preserves the phase space measure.  The coordinate $\rho = x - X(t)$ is co-moving with the soliton.  The Hamiltonian, when expressed in the new variables, is nonlocal in space and possesses an infinite set of higher order vertices for the fluctuation field, depending on the background soliton solution \cite{Dashen:1974ci,Dashen:1974cj,Goldstone:1974gf,Gervais:1974dc,Callan:1975yy,Christ:1975wt,Gervais:1975pa,Tomboulis:1975gf}.  The fluctuations can be expanded in creation and annihilation operators, and the vacua of the one-soliton sector, which are annihilated by all of the annihilation operators and labeled by the eigenvalues of $P$, are the one-soliton states of the theory.

In order to make use of the soliton-sector Hamiltonian for practical computations, one typically employs two approximations: the semiclassical approximation and the adiabatic approximation.  The semiclassical approximation is the usual small coupling expansion, where the coupling, $g$,  is a parameter in the potential controlling cubic and higher order interaction terms.  By a scaling argument, $g^2$ can be identified with $\hbar$.  The adiabatic approximation, meanwhile, treats time derivatives of the soliton collective coordinate as small.  It is the implementation in quantum field theory of Manton's approximation for time-dependent soliton solutions \cite{Manton:1981mp} of classical field theory. In fact, it is common to tie these two approximations together by assigning a particular $g$ scaling to time derivatives, $\partial_t \sim O(g)$.\footnote{There is good reason to do so.  The static-soliton profile with a time-dependent collective coordinate does not solve the time-dependent field theory equations of motion.  Keeping $\partial_t \sim O(g)$ ensures that the resulting tadpole for the quantum fluctuation field can be grouped with the interaction Hamiltonian and treated perturbatively.}

One can formally define the soliton effective Hamiltonian $H_{\rm eff}[P]$ by path integrating over the field-theoretic fluctuations. In principle the Hamiltonian can then be computed perturbatively in both expansion parameters.  At lowest order in both one recovers the standard relativistic energy $\sqrt{P^2 + M_{0}^2}$, where $M_0$ is the classical soliton mass.\footnote{This is the result in the simplest two-dimensional models where the only soliton collective coordinate is the position degree of freedom, $X$.}  In this language the adiabatic expansion is a small momentum \emph{transfer} expansion, $\dot{P} \ll 1$, while the small $g$ expansion provides quantum field-theoretic corrections.   

With a few notable exceptions \cite{Drukier:1981fq,Bachas:1992dw,Demidov:2015nea}, and excluding integrable theories where other techniques are available, almost all work on solitons in quantum field theory has been in the adiabatic or small momentum transfer limit.  For example, in supersymmetric theories where the solitons are BPS states one can sometimes use nonperturbative techniques in $g$ to recover the quantum-exact mass $M(g) = M_0 + O(g^2)$.  Classic references include \cite{Cecotti:1992rm,Seiberg:1994rs}, and a recent review may be found in \cite{MR3307344}.  However, very little is known about solitons in non-integrable theories in the opposite limit of high momentum transfer but small $g$.

Understanding the behavior of solitons in the high momentum transfer regime, $\Delta P \sim M_{0}$, is extremely important for certain foundational questions in quantum field theory.  For example, should one consider soliton-antisoliton virtual pairs running in loops when one computes quantum corrections to ordinary processes involving perturbative particles?  Naively the answer is yes, since the optical theorem instructs one to sum over all intermediate states.  Less naively the answer is no, since arguments based on a coherent-state picture of the soliton-antisoliton pair indicate such contributions will be exponentially suppressed in the coupling and hence beyond the regime of applicability of the asymptotic series in $g$, which perturbation theory provides for any given observable.  

However, other arguments \cite{Banks:2012dp,Papageorgakis:2014dma} suggest that the exponential suppression is governed not by the coupling $g^2$ \emph{per se}, but by the ratio of the soliton's Compton wavelength to its size.  For many theories this ratio is essentially the same as $g^2$ but there are notable exceptions such as instanton-solitons in five-dimensional gauge theories and small black holes in supergravity.  In such cases one should ask: Is it possible that the contributions of small-sized solitons running in loops are important for perturbative processes?  Recent computations suggest that this is indeed the case \cite{Bossard:2015foa,Bossard:2017kfv}.  

The idea of \cite{Banks:2012dp} is to employ crossing symmetry to relate the creation of a virtual soliton-antisoliton pair to an (off-shell) process in the one-soliton sector of the theory in which a soliton absorbs or emits a high-momentum perturbative particle.\footnote{A cautionary remark is in order.  Crossing symmetry is a symmetry of the exact quantum field theory.  Due to the possibility of Stokes phenomena, the semiclassical limit might not commute with the analytic continuation to the crossed channel.}  The amplitude for the latter process is captured by a form factor---that is, a matrix element of the scalar field between soliton states, $\llangle P_f | \phi | P_i \rrangle$. Reference \cite{Papageorgakis:2014dma} improved on previous work by making the Lorentz invariance of the form factor manifest, a result achieved by working with the boosted-soliton profile and the relativistic effective Hamiltonian, $H_{\rm eff}[P]$.  However the computation in \cite{Papageorgakis:2014dma} still assumed a small momentum transfer compared to the soliton mass, thus leaving speculation about the regime of high momentum transfer open.\footnote{Indeed, the authors of \cite{Papageorgakis:2014dma} did not fully realize this in the earlier versions of the work. Clarifying this point has been a significant motivation for the current paper.  The authors of \cite{Papageorgakis:2014dma} thank Sergei Demidov, Dmitry Levkov and Edward Witten for early related discussions.}

In this paper we demonstrate that one can access the high momentum transfer regime of solitons in two-dimensional linear sigma models, by working directly with the exact field-theoretic soliton-sector Hamiltonian obtained in \cite{Gervais:1975pa,Tomboulis:1975gf}.  By carrying out a saddle point approximation of the path integral over $(\varphi,\varpi)$ in $g$---but working exactly with the nonlocal terms---we will obtain a saddle point equation for the fluctuation field $\varphi$ in the soliton sector.  With $V(\phi)$ the field theory potential, this reads as
\begin{equation}\label{fse}
(\partial_t - \beta[P,\varphi] \partial_\rho)^2 \varphi - \partial_{\rho}^2 \varphi + \frac{dV}{d\phi} \bigg|_{\varphi} =  f[\dot{P},\varphi]~,
\end{equation}
where the generalized velocity $\beta$ and forcing term $f$ are functionals of the soliton momentum $P$ and the field $\varphi$:
\begin{equation}
\beta[P,\varphi] := \frac{P + \int d\rho \dot{\varphi} \varphi'}{\int d\rho \varphi^{\prime 2}} ~, \qquad f[\dot{P},\varphi] := -\frac{\dot{P} \uppsi_0}{\int d\rho \uppsi_0 \varphi'} ~.
\end{equation}
Here $\uppsi_0 = \frac{1}{\sqrt{M_0}} \phi_{0}'$ is the normalized zero-mode fluctuation around the static soliton solution, $\phi_0(\rho)$, while dots and primes are used to denote derivatives with respect to $t$ and $\rho$ respectively.  In the limit of zero momentum transfer, so that $P$ is constant, the forcing term vanishes, and this integro-differential equation reduces to one obtained already in \cite{Gervais:1975pa}.  It is solved by the boosted soliton profile: $\varphi = \phi_0 (\rho/\sqrt{1-\beta^2})$, with $\beta$ the soliton velocity related to the momentum via $P = M_0 \beta /\sqrt{1-\beta^2}$.

In \eqref{fse}, which we will refer to as the \emph{forced soliton equation}, $P(t)$ should be viewed as a given function of time, and therefore the solution $\varphi$ is a functional of $P$.  Inserting this solution back into the field-theoretic soliton-sector Hamiltonian then yields $H_{\rm eff}[P]$ at tree-level in the coupling $g$, but to all orders in the time-derivative expansion. A solution $\varphi$ to \eqref{fse} can be thought of as a nonlinear soliton analog of the Li\'enard--Wiechert potential in electromagnetism. A key difference is that the soliton degree of freedom $X(t)$ is not external to the full theory but is governed by the effective Hamiltonian $H_{\rm eff}[P]$.

Quantum fluctuations around a solution to \eqref{fse} can be treated in the usual perturbative manner.\footnote{One degree of freedom associated with the zero-mode must be eliminated from the fluctuation field $\delta\varphi$.  There are standard techniques for doing so, as we will review.}  Integrating out these degrees of freedom results in the one-loop and higher order contributions to the soliton effective Hamiltonian, viewed as an expansion in $g$.  In this paper we restrict ourselves to the one-loop analysis.  We manage to carry out the relevant path integral over fluctuations in closed form, giving the result in terms of classical data associated to the quadratic fluctuation Hamiltonian.  However we eventually must restrict to the case of constant $P$, where the saddle-point solution to \eqref{fse} is known, to carry the computation to completion.  

Even in this case, we find a rather nontrivial quadratic Hamiltonian to diagonalize for the fluctuations.  Doing so requires an extension to an approach presented in the appendix of the classic paper by Christ and Lee \cite{Christ:1975wt}, and we couch the analysis in the language of Williamson's Theorem \cite{MR1507138}.  One output, which will be useful for higher order perturbative computations, is the explicit form of the normal-mode fluctuations around the boosted soliton; these satisfy an orthogonality condition with respect to the zero-mode of the static soliton.

Finally, we use the soliton effective Hamiltonian to define \emph{semiclassical soliton form factors}.  These capture the leading-in-$g$ behavior of the corresponding soliton form factors and are valid at arbitrary momentum transfer.  We are able to carry out the quantum mechanical path integral for these semiclassical form factors and express the result in terms of a generating functional.  The generating functional uses the soliton effective Hamiltonian, evaluated on a time-dependent solution to the forced soliton equation determined by a source.  The differential operator that acts on the generating functional to produce the semiclassical soliton form factor is constructed from a constant $P$ background solution and is thus known in terms of the classical soliton profile.  We demonstrate that our formula reduces to the expected result, in terms of a Fourier transform of the classical profile, in the low momentum transfer regime.

The rest of this paper is organized as follows.  In Section \ref{sec:CT} we recall the canonical transformation of the phase-space path integral that separates the soliton collective coordinate and its conjugate momentum from the remaining set of field-theoretic degrees of freedom, and we recall the form-factor computations that motivated this investigation.  The resulting soliton-sector Hamiltonian is analyzed in Section \ref{sec:SP}, where it is shown that \eqref{fse} arises as the saddle-point equation for the fluctuation field around an accelerating soliton with phase-space trajectory $(X(t),P(t))$.  In Section \ref{sec:SHO} we set up and evaluate the saddle-point approximation for the path integral over the field-theoretic fluctuations around a solution to \eqref{fse}.  In Section \ref{sec:FW} we apply our machinery to semiclassical soliton form factors.  An investigation of solutions to \eqref{fse} beyond small momentum transfer is left to future work, and we outline some potential approaches at the end of Section \ref{sec:FW}.

\section{Canonical Transformation to the One-soliton Sector}\label{sec:CT}

We begin our discussion by setting up notation and conventions, as well as briefly reviewing some of the necessary background material.

\subsection{Preliminaries}

We consider the class of 2D linear sigma models with classical action
\begin{equation}\label{eq:start}
S = \int d^2 x \left( -\half \pd_\mu \phi \pd^\mu \phi - V_0(m_0;\phi) \right)~.
\end{equation}
We assume that the minima of $V_0(m_0;\phi)$ are gapped and associated to a spontaneously broken discrete symmetry.  The parameter $m_0$ controls the mass gap to the perturbative spectrum.  Spacetime points are labeled by $x^\mu = (t,x)$, and we work in signature $(-,+)$.  

If $M_{\rm vac} := \{ \phi ~|~ V_0(m_0; \phi) = V_{\rm min} \}$ has multiple components then there exist classical solitons called kinks.  These are finite-energy time-independent solutions to the equation of motion,
\begin{equation}\label{classicaleom}
\frac{\pd^2 \phi}{\pd x^2} - \frac{dV_0(m_0;\phi)}{d\phi} = 0~,
\end{equation}
that asymptote to different vacua as $x \to \pm \infty$.  We denote such a solution by
\begin{equation}\label{classicalsoliton}
\phi = \phi_0(m_0;x - X)~.
\end{equation}
The free parameter $X$ is the center-of-mass position of the kink. Prototypical examples within this class, along with their static classical solutions, are \emph{c.f.} \cite{Rajaraman:1982is}
\begin{align}\label{models}
\textrm{$\phi^4$ theory:} \qquad &  V_0 = \frac{1}{g^2} \left( g^2 \phi^2 - \frac{1}{4} m_{0}^2 \right)^2~, \qquad &   \phi_0 = \frac{m_0 }{2 g}\tanh\left( \frac{m_0}{\sqrt 2} (x- X)\right)~, \cr
\textrm{sine-Gordon:} \qquad &  V_0 = \frac{m_{0}^2}{g^2} \left( 1 - \cos{(g \phi)} \right)~,  \qquad & \phi_0 = \frac{4}{g}\arctan\left( e^{m_0 (x- X)} \right)~.
\end{align}
In the $\phi^4$ model there are two classical vacua at $\phi = \pm m_{0}/2g$, while in the sine-Gordon model there is an infinite sequence at $\phi = \frac{2\pi n}{g}$, $n \in \mathbbm{Z}$.  Expanding around these vacua, one finds that the tree-level masses of the fundamental particles, the ``mesons,'' are $\sqrt{2} m_0$ and $m_0$ in the $\phi^4$ and sine-Gordon models respectively.

Note that both of the above potentials have the scaling property $V_0(m_0;\phi) = \frac{1}{g^2} \widetilde{V}_0 (m_0;\widetilde{\phi})$, where the function $ \widetilde V_0(m_0; x)$ does not depend on $g$, and $\widetilde{\phi} = g\phi$. It is common in the soliton literature \cite{Goldstone:1974gf, Jackiw:1977yn, Rajaraman:1982is} to make the additional assumption that the potential is of this form, in which case, the action can be written
\begin{equation}\label{eq:rescaledS}
S = \frac{1}{g^2} \int d^2 x \left( -\half \pd_\mu \widetilde{\phi} \pd^\mu \widetilde{\phi} -  \widetilde V_0(m_0; \widetilde{\phi}) \right)~.
\end{equation}
Thus the $g$-expansion can be thought of as the (semiclassical) $\hbar$-expansion.  We assume the potentials we work with in this paper have this scaling property.
 
In the quantum theory, bare and renormalized parameters must be related through appropriate counterterms.  Perturbative sector computations in these models reveal logarithmic divergences only, which can be eliminated through mass renormalization.  The coefficient of the mass counterterm, $\Delta m^2$, can be computed order by order in perturbation theory once a renormalization prescription is given, and the one-loop contribution, $\delta m^2$, participates in the evaluation of the one-loop correction to the soliton effective Hamiltonian.  

A standard renormalization prescription can be made for the class of linear sigma models discussed above where the effect of the counterterms is such that the renormalized potential, $V(\phi) = V_0(\phi) + V_{\rm c.t.}(\phi)$, has the same form as $V_0$, but with $m_{0}^2$ replaced with $m^2 + \Delta m^2$, where $m^2$ is a finite renormalized mass, and the condition fixing $\Delta m^2$ is that the tadpole for the fluctuation field around the vacuum vanishes to all orders in perturbation theory.  See \cite{Rebhan:1997iv} for a fuller discussion.  In the case of $\phi^4$ theory, for example, this condition implies that the quantum vacua are at $\llangle \phi \rrangle = \pm m/2g$. Using this condition, we will write the renormalized potential as
\begin{align}\label{Vfull}
V(\phi) :=&~ V_0(m_0; \phi) + V_{\rm c.t.}(\phi) \cr
=&~ V_0(m;\phi) + V_{\Delta m^2}(\phi) \cr
=&~ V_0(m;\phi) + V_{\delta m^2}(\phi) + \cdots ~,
\end{align}
where the ellipsis denotes two-loop and higher contributions.  Thus the relevant background configuration for a perturbative analysis of the quantum kink is
\begin{equation}
\phi = \phi_0(m;x - X)~,
\end{equation}
This is the classical kink solution where the bare mass parameter, $m_0$, has been replaced with the renormalized parameter, $m$.  Henceforth, any appearance of $V_0(\phi)$ or $\phi_0(x -X)$, without the mass parameter made explicit, will refer to $V_0(m;\phi)$ and $\phi_0(m,x-X)$.         

\subsection{Soliton States}

Soliton states are elements of the one-particle Hilbert space and are labeled by momentum $P$ \cite{Goldstone:1974gf}.  They carry a conserved topological charge, associated with the current $J_\mu = \epsilon_{\mu\nu} \pd^\nu \phi$, and are orthogonal to the perturbative-sector states.  They are defined through a three-step process:

\paragraph{Step 1:} One begins with the renormalized Hamiltonian arising from \eqref{eq:start}, corrected by perturbative sector counterterms, and given by
\begin{equation}\label{HamiltonianOld}
H = \int dx \left\{ \frac{1}{2} \pi^2 +  \half (\pd_x \phi)^2 + V(\phi) \right\}~.
\end{equation}
In terms of this Hamiltonian, the phase-space path integral for the transition amplitude takes the form
\begin{equation}\label{genfunnew}
\mathcal{Z}_{fi} = \int [D\phi D\pi] \Psi_{f}[\phi(t_f, x)]^\ast \Psi_i[\phi(t_i,x)] \exp \bigg\{ i \int_{t_i}^{t_f} dt \left[ \int dx \dot{\phi} \pi  - H \right] \bigg\}~.
\end{equation}
Here $\Psi_{i,f}[\phi]$ are wave functionals for the initial and final state of the field at times $t_i$ and $t_f = t_i + T$, and $\ZZ_{fi} \equiv \llangle \Psi_f | \ZZ |\Psi_i \rrangle$ is the matrix element of the evolution operator.

\paragraph{Step 2:} Next, a canonical transformation on (infinite-dimensional) phase space $(\phi ;\pi) \mapsto (X,\chi; P,\varpi)$ is performed by considering the coordinate transformation
\begin{equation}\label{cxfmphi}
\phi(t,x) = \phi_{0}(x-X(t)) + \chi(t,x-X(t))~.
\end{equation}
The modulus $X$ has been promoted to a dynamical variable (a \emph{collective coordinate}) and $\chi$ represents field fluctuations around the soliton.  Then one makes the following ansatz for the momentum variables :
\begin{equation}\label{cxfmpi}
\pi(t,x) = \pi_0[X,\chi; P,\varpi] \pd_x \phi_0(x-X(t)) + \varpi(t,x-X(t))~.
\end{equation}
In order to preserve the number of degrees of freedom, constraints must be imposed on the new fields $\chi,\varpi$.  Since $\pd_x \phi_0$ solves the linearized equation of motion around the soliton solution, the constraints are chosen to eliminate this zero-frequency degree of freedom from the new fields
\begin{equation}\label{constraints1}
\int dx \chi \pd_x \phi_0 = 0~, \qquad \int dx \varpi \pd_x \phi_0 = 0~.
\end{equation}

The quantity $\pi_0$ is determined by demanding that the transformation be canonical---\ie\, by requiring that the standard Poisson bracket of $\phi,\pi$ implies $\{X,P\} = 1$ and the standard Dirac bracket for $\varphi,\varpi$, and vice versa.  This leads to \cite{Tomboulis:1975gf,Gervais:1975pa}
\begin{align}
  \label{eq:1}
  \pi_0 = -\frac{P + \int d \rho \varpi  \chi' }{M_0[1 + (1/M_0) \int d\rho \phi_{0}' \chi']} \;,
\end{align}
with
\begin{align}\label{massform}
  M_0 = \int d\rho \phi_{0}^{\prime\, 2} 
\end{align}
the classical soliton mass.  Here we have introduced the kink-comoving coordinate $\rho = x - X(t)$.  Derivatives with respect to $\rho$ are denoted with a prime.

As integrals over $\rho$, like those in \eqref{eq:1}, will appear quite often, we will sometimes employ a bra-ket notation:\footnote{To minimize confusion we use a double bra-ket for quantum field theory inner products.}
\begin{equation}
\langle f | g \rangle := \int_{-\infty}^{\infty} d\rho f(t,\rho)^\ast g(t,\rho)~.
\end{equation}
Thus $M_0 = \langle \phi_{0}' | \phi_{0}' \rangle$, and \eqref{eq:1} can also be expressed more compactly as
\begin{equation}\label{pi0result}
\pi_0 = - \frac{P + \langle \varpi | \chi' \rangle}{\langle \phi_{0}' | \phi_0' + \chi' \rangle } ~.
\end{equation}

In terms of the new variables, \eqref{cxfmphi} and \eqref{cxfmpi}, the Hamiltonian \eqref{HamiltonianOld} is
\begin{equation}\label{Ham}
H = M_0 + \frac{M_0 \left( P + \langle \varpi | \chi' \rangle \right)^2}{2 \langle \phi_{0}' | \phi_{0}' +\chi' \rangle^2} + \int d\rho \left\{ \half \varpi^2 + \half {\chi'}^2 + \sum_{n \geq 2} \frac{1}{n!}  V_{0}^{(n)}(\phi_0) \chi^n + V_{\Delta m^2}(\phi_0 + \chi) \right\},
\end{equation}
where $ V^{(n)}$ denotes the $n^{th}$ derivative of the potential with respect to $\phi$.  We will refer to this as the soliton-sector Hamiltonian.  It is important to note that the canonical transformation \eqref{cxfmphi}, \eqref{cxfmpi}, with \eqref{eq:1}, is a transformation of classical phase-space variables.  In the canonical formalism, though, it is straightforward to extend it to a transformation of operators that preserves the quantum commutator \cite{Tomboulis:1975gf}.  This requires a choice of operator ordering for \eqref{cxfmpi}.  Upon choosing the Weyl prescription, for example, one is led to an additional term in the Hamiltonian beyond \eqref{Ham}.\footnote{See also \cite{Papageorgakis:2014jia} for a discussion in the context multi-component scalar theories.}  This ``quantum potential'' can also be obtained from the path integral formalism through a careful treatment starting from the discretized definition, (where Weyl ordering corresponds to the midpoint prescription for the momenta) \cite{Gervais:1976ws}.  In terms of scaling in $g$, the quantum potential is an $O(g^2)$ correction, or two-loop effect, and hence will not be relevant for us in this paper.  However,  the quantum potential should be understood to be included in any expression appearing below that utilizes the exact soliton-sector Hamiltonian.

In order to write the transition amplitude \eqref{genfunnew} in the new variables we must give a precise description of how the constraints \eqref{constraints1} are to be implemented in the path integral.  For this purpose it is useful to introduce an orthonormal basis of modes for the fluctuations around the classical static solution, $\phi_0(\rho)$.  These modes solve the eigenproblem
\begin{align}
  \label{eq:3}
  \Big( - \frac{\pd^2}{\pd\rho^2} + \frac{d^2 V_{0}}{d \phi^2} \bigg|_{\phi_0} \Big) \uppsi_n(\rho) = \omega_n^2 \uppsi_n(\rho)\; ,
\end{align}
which arises from a linearization of \eqref{classicaleom} around the static soliton solution, \eqref{classicalsoliton}, with the replacement $m_0 \to m$ as explained under \eqref{Vfull}.  The modes $\{ \uppsi_n \}$ are known explicitly for many field theory potentials of interest, but we will not need their detailed form; we only wish to emphasize a few key points that hold for the class of models we consider.  The spectrum is positive, and there is a unique zero-mode given by
\begin{equation}\label{eq:zeromode}
\uppsi_0 = \frac{1}{\sqrt{M_0}} \phi_{0}' ~.
\end{equation}
In terms of this zero-mode the constraints take the form
\begin{equation}\label{eq:2}
\langle \uppsi_0 | \chi \rangle = 0~, \qquad \langle \uppsi_0 | \varpi \rangle = 0~.
\end{equation}
Depending on the details of the potential there might, or might not, exist additional discrete $L^2$-normalizable modes corresponding to excited states of the kink.  These will be followed by a continuum for the theory defined on $\rho \in (-\infty,\infty)$, representing perturbative particle states in the presence of the kink.  

For later purposes---especially the one-loop computation of Section \ref{sec:SHO}---it will be essential to regularize the theory by putting it in a box of size $L$ with appropriate boundary conditions imposed at $\rho = \pm L/2$, so that the spectrum of fluctuations around the kink is discrete.  We make a brief digression here to explain this in some detail, since having the basic framework in place now will prove convenient later.

 The boundary conditions at $\rho = \pm L/2$ must ensure that the operator in \eqref{eq:3} is Hermitian and should maintain a well-defined variational principle for \eqref{eq:start}.  In subsection \ref{sec:oneloopcp} where we review the one-loop computation of \cite{Dashen:1974cj,Jain:1990dq}, we will take periodic boundary conditions and employ mode number regularization as in the original works.  Other choices are possible and yield the same results provided that the regulators in the soliton and perturbative sectors are chosen consistently.\footnote{The questions of boundary conditions and regularization are subtler for supersymmetric kink solitons in theories with fermions, and this led to a flurry of activity on the subject in the late '90's and early 2000's.  See \eg\,\cite{Goldhaber:2004kn} for a review with references.  In contrast, for the simple bosonic models of \eqref{models}, the original approach and results of Dashen, Hasslacher, and Neveu \cite{Dashen:1974cj} have been validated many times.}  This point has been nicely emphasized in recent work by Evslin \cite{Evslin:2020rgm}. 

As long as the size of the box is taken much larger than any length scale in the potential, the square-normalizable bound states of \eqref{eq:3} will continue to exist with box eigenvalues, $\omega_{n}(L)^2$, and box wavefunctions that differ from those of the theory on $\mathbbm{R}$ by corrections in $L$ that are exponentially small at large $L$.  Furthermore, although the spectrum is made discrete by the box regularization, one can still distinguish those eigenfunctions corresponding to bound states in the theory on $\mathbbm{R}$ from those eigenfunctions corresponding to scattering states by the value of $\omega_{n}(L)^2$, at least when $L$ is large enough.  Specifically, if $\omega_{n}(L)^2 < \frac{d^2 V}{d\phi^2}(\phi_0(\rho))$ for all $|\rho| > L/2$, then $\uppsi_n$ corresponds to a bound state of the potential while those modes with $\omega_{n}(L)^2 \geq \frac{d^2 V}{d\phi^2}(\phi_0(\rho))$ for all $|\rho| > L/2$ correspond to scattering states.  Strictly speaking, the previous comment assumes that $\frac{d^2 V}{d\phi^2}(\phi_0(\rho))$ is a symmetric function of $\rho$.  If it is not, one can easily modify the statement to account for the different behaviors of $\frac{d^2 V}{d\phi^2}(\phi_0(\rho))$ at $\rho = \pm L/2$. We thus have the orthonormality and completeness relations for the theory in the box:
\begin{equation}
\langle \uppsi_m | \uppsi_n \rangle = \delta_{mn} ~, \qquad \mathbbm{1} = \sum_{n} |\uppsi_n \rangle \langle \uppsi_n|~.
\end{equation}
Our conventions are that $n = 0,1,\ldots,n_{\rm b}-1$ correspond to the bound states of $-\pd_{\rho}^2 + \frac{d^2 V}{d\phi^2} |_{\phi_0}$ on $\mathbbm{R}$ with $n = 0$ the zero-mode, while $n = n_{\rm b}, n_{\rm b} + 1,\ldots$ correspond to the scattering states.

In the box, the equality \eqref{eq:zeromode} is no longer true.  The relationship given there will receive corrections that are exponentially small in $L$ at large $L$.  Furthermore, the eigenvalue $\omega_{0}^2$ will not be zero.  Our goal is to study the theory on $\mathbbm{R}$, and we are only using $L$ as a regulator.  Thus we have some freedom in how we choose to define the transformation to the soliton sector when $L$ is finite.  Rather than using $\phi_{0}'$ for the constraints, we will use $\uppsi_0$.  Again, these agree when $L \to \infty$.  However the choice $\uppsi_0$ seems more appropriate at finite $L$ since this way we are exchanging the lowest energy eigenmode around the static kink for a collective coordinate.  The phase space coordinate transformation will be canonical with the $\uppsi_0$ constraints at finite $L$ provided \eqref{cxfmpi} and \eqref{pi0result} are written as
\begin{equation}\label{newpicx}
\pi(t,x) = - \left( \frac{P + \langle \varpi | \phi_{0}' + \chi' \rangle }{\langle \uppsi_0 | \phi_{0}' + \chi' \rangle } \right) \uppsi_0(x - X(t)) + \varpi(t,x - X(t))~.
\end{equation}
Now the soliton sector Hamiltonian is 
\begin{equation}\label{HamL}
H =  \frac{ (P + \langle \varpi | \phi_{0}' + \chi' \rangle )^2}{2 \langle \uppsi_0 | \phi_0' + \chi' \rangle^2} + \int_{-L/2}^{L/2} d\rho \left\{ \ff{1}{2} \varpi^2 + \ff{1}{2} (\phi_{0}' + \chi' )^2 + V(\phi_0 + \chi) \right\}~,
\end{equation}
and it is equivalent to \eqref{Ham} in the $L \to \infty$ limit.

Returning to the main thread of the discussion, then, we can write mode expansions
\begin{align}
  \label{eq:4}
& \chi(t,\rho) = \sum_n \chi_n(t) \uppsi_n(\rho)~, \qquad  \varpi(t,\rho) = \sum_n \pi_n(t) \uppsi_n(\rho) ~,
\end{align}
and the constraints \eqref{eq:2} set $\chi_0(t),\pi_0(t)$ to zero.  It will be convenient for us to work with a real basis $\{ \uppsi_n \}$~, and therefore the modes $\chi_n,\pi_n$ are real-valued.  We emphasize that the basis $\{ \uppsi_n \}$ does \emph{not} diagonalize the quadratic Hamiltonian in $\chi,\varpi$ unless $P=0$.

With the soliton sector Hamiltonian, \eqref{HamL}, the transition amplitude \eqref{genfunnew} is expressed as an integral over the new variables with delta functionals enforcing the constraints:
\begin{align}\label{Zfi}
\mathcal{Z}_{fi} =&~ \int [DXDP] \int [D\chi D\varpi] \delta(\langle \uppsi_0 |\chi \rangle ) \delta(\langle \uppsi_0 | \varpi \rangle) \Psi_f[X_f,\chi(t_f,x)]^\ast \Psi_i[X_i, \chi(t_i,x)] \times \cr
&~ \qquad \qquad \qquad \times \exp\left\{ i \int_{t_i}^{t_f} dt \left( P \dot{X} + \langle \varpi | \dot{\chi} \rangle - H \right) \right\} \cr
=&~  \int [DXDP] \int [D\chi D\varpi D\lambda D\nu]  \Psi_f[X_f,\chi(t_f,x)]^\ast \Psi_i[X_i, \chi(t_i,x)]  \times  \cr
&~ \qquad \qquad \qquad \times \exp\left\{ i \int_{t_i}^{t_f} dt \left( P \dot{X} + \langle \varpi | \dot{\chi} \rangle - \lambda \langle \uppsi_0 | \chi \rangle - \nu \langle \uppsi_0 | \varpi \rangle - H \right) \right\} ~.  \raisetag{24pt}
\end{align}
In the second form of the expression the delta functionals have been represented by functional integration over Lagrange multipliers $\lambda(t),\nu(t)$.  We denote by
\begin{equation}\label{ssH}
H_{T} = H + \lambda \langle \uppsi_0 | \chi \rangle + \nu \langle \uppsi_0 | \varpi \rangle~
\end{equation}
the total Hamiltonian, which includes the Lagrange multipliers.

\paragraph{Step 3:}  Now that the transition amplitude has been expressed in appropriate variables, we can define the soliton states.  A soliton state of momentum $P$ has the form $|\Psi_P \rrangle =  | P \rrangle \otimes |\Psi_{0,P} \rrangle$ with position-basis wavefunctional
\begin{equation}
\llangle X, \chi  | \Psi_P \rrangle = \Psi_P[X,\chi] = \frac{1}{\sqrt{2\pi}} e^{i P X} \Psi_{0,P}[\chi] ~,
\end{equation}
where $\Psi_{0,P}$ is the normalized ground-state wavefunctional of the $\chi$-$\varpi$ theory---that is the theory defined by the Hamiltonian \eqref{ssH}, where $P$ is treated as a (generally time-dependent) background parameter.  The notation $\Psi_{0,P}$ is meant to emphasize the dependence of this ground-state wavefunctional on $P$, but we will often omit the $P$ subscript for brevity.

The wavefunctional $\Psi_0$ can be computed perturbatively in the semiclassical expansion.  If $\chi =\overbar{\chi} + \delta\chi$, where $\overbar{\chi}$ is a solution to the classical equations of motion following from \eqref{ssH} and $\delta\chi$ is the fluctuation field, then at leading (one-loop) order the wavefunctional takes the form of a Gaussian in the fluctuation field $\delta\chi$.  If the fluctuation field is written in terms of creation/annihilation operators $\ah^\dag$ and $\ah$ that diagonalize the quadratic part of the Hamiltonian, then the ground state $|\Psi_{0} \rrangle$ is the state annihilated by all of the annihilation operators.

The ground-state wavefunctional is used to define the soliton effective Hamiltonian, $H_{\rm eff}[P]$, via
\begin{align}\label{Heffdef}
e^{- i \int dt H_{\rm eff}[P]} :=&~  \int [D\chi D\varpi D\lambda D\nu]  \Psi_{0}[\chi(t_f,x)]^\ast \Psi_{0}[\chi(t_i,x)] e^{ i \int dt \left( \langle \varpi | \dot{\chi} \rangle - H_T \right)} ~.
\end{align}

The main goal of this paper is to construct the saddle-point approximation to $H_{\rm eff}[P]$ for general time-dependent $P(t)$; we will present these results in Section~\ref{sec:SP}.  In Section \ref{sec:SHO} we will construct the one-loop $\Psi_{0}$ in terms of a symplectic transformation that diagonalizes the part of $H_T$ that is quadratic in fluctuations.\footnote{The wavefunctional for the static soliton, $\Psi_{0,P=0}$, has been discussed recently in \cite{Evslin:2020qow}, where it was obtained by acting on the perturbative sector vacuum with an appropriate displacement operator.}  We then use the wavefunctional to complete the saddle-point computation of \eqref{Heffdef}.  The construction is fully explicit in the case of constant $P$.   Before we turn to that technical analysis, we will describe an important physical application where $H_{\rm eff}[P]$ will be useful.

\subsection{Motivation from Soliton Form Factors}\label{sec:solFF}

One of the main motivations behind this work is the study of soliton form factors to leading order in the perturbative expansion and beyond the regime of small momentum transfer.  Soliton form factors are simply matrix elements of operators between soliton states: $\llangle \Psi_{P_f} | \widehat{\OO} | \Psi_{P_i} \rrangle$.  For example, one of the most basic and important form factors is the matrix element of the original scalar field, $\widehat{\OO} = \widehat{\phi}(t,x)$.  By crossing symmetry, this form factor determines the amplitude for a perturbative particle to create a soliton-antisoliton pair \cite{Coleman:1975ia,Banks:2012dp,Papageorgakis:2014dma}.

At leading order in the semiclassical approximation, any polynomial in the original fields $\phi,\pi$ restricts to a function of $X$ and $P$ obtained by evaluating \eqref{cxfmphi}, \eqref{cxfmpi} on the saddle-point solution:
\begin{equation}
\OO[\pi,\phi] = \OO[\pibar[P,X],\phibar[P,X]] \left(1 + O(g) \right)~,
\end{equation}
where
\begin{equation}
\phibar = \phi_0(x - X) + \overbar{\chi}(t,x-X)~, \qquad \pibar = \pi_0[X,\overbar{\chi};P,\varpibar] \pd_x \phi_0 (x - X) + \varpibar(t,x-X)~,
\end{equation}
with $(\overbar{\chi},\varpibar)$ a solution to the classical equations of motion stemming from the soliton-sector Hamiltonian, \eqref{ssH}.  It follows that the leading semiclassical approximation to the soliton form factor reduces to a matrix element in the collective-coordinate quantum mechanics\footnote{Corrections can be computed perturbatively in the $\chi$-$\varpi$ field theory.  See e.g.~\cite{Gervais:1975pa}.} with Hamiltonian $H_{\rm eff}[P]$:
\begin{align}\label{spFF}
\llangle \Psi_{P_f} | \widehat{\OO} | \Psi_{P_i} \rrangle =&~ \int [DXDP] \int [D\chi D\varpi D\lambda D\nu] \Psi_{P_f}^\ast \Psi_{P_i} e^{i \int dt \left( P \dot{X} + \langle \varpi | \dot{\chi} \rangle - H_T \right)} \OO[\pi,\phi] \cr
=&~ \int [DX DP] \frac{1}{2\pi} e^{ i (P_i X_i - P_f X_f)} e^{i \int dt (P \dot{X} - H_{\rm eff}[P])} \OO[\pibar,\phibar] \left( 1 + O(g) \right)~.
\end{align}
As we will see, $H_{\rm eff}[P]$ has an expansion of the form 
\begin{equation}\label{Heffexpansion}
H_{\rm eff} = H_{\rm eff}^{(-2)} + H_{\rm eff}^{(0)} + O(g) ~,
\end{equation}
where the superscripts indicate the order in $g$.  The first (tree-level) term arises from evaluating $H_T$, given in \eqref{ssH}, on the background solution $\overbar{\chi},\varpibar$, while the second (one-loop) term comes from integrating out the fluctuations around this solution via saddle point approximation.  It follows from \eqref{spFF}  that it is sufficient to keep only these first two terms in order to capture the leading semiclassical behavior of soliton form factors.

This discussion also allows us to highlight when and why it is important to go beyond the usual adiabatic/small momentum-transfer limit.  First, if there is no $X$-dependent insertion in \eqref{spFF}, then the soliton momentum is conserved.  This can be seen explicitly from \eqref{spFF} by first carrying out the path integral over $X$, resulting in a delta functional setting $P$ equal to a constant.  The overall matrix element then carries a factor of $\delta(P_f - P_i)$.  For example, in the case where there is no insertion one has
\begin{equation}\label{identityFF}
\llangle \Psi_{P_f} | \Psi_{P_i} \rrangle = \delta(P_f - P_i) e^{-i H_{\rm eff}[P_f] T} = \delta(P_f - P_i) e^{- i H_{\rm eff}[P_f] t_f + i H_{\rm eff}[P_i] t_i}~,
\end{equation}
where $T = t_f - t_i$, and $H_{\rm eff}[P_f]$ can be evaluated perturbatively in $g$ under the restriction that $P_f$ is constant.  The Lorentz invariance of the theory dictates that
\begin{equation}
H_{\rm eff}[P] = \sqrt{P^2 + M^2}~, \qquad (\textrm{constant $P$})~,
\end{equation}
where $M = M_0 (1 + O(g^2))$ is the exact quantum mass of the soliton.  The verification of this relativistic energy to one-loop accuracy (in non-integrable models) is a classic result going back to \cite{Christ:1975wt,Jain:1990dq}. However, as far as we are aware it has not been demonstrated directly from the path integral formalism of \cite{Gervais:1975pa}, as we will do in Section \ref{sec:SHO}.  

Ultimately, we are interested in matrix elements of operators that do carry $X$-dependence, which means that the soliton momentum is not conserved.  Let us return to the example of the scalar-field form factor, $\llangle \Psi_{P_f} | \widehat{\phi} | \Psi_{P_i} \rrangle$.  If one works to leading order in the derivative expansion of $H_{\rm eff}$ (by treating $P$ as constant),  then to leading order in both the derivative and semiclassical expansions this form factor is given by the Fourier transform of the classical soliton solution, $\phi_0$, expressed as a function of $k = P_f - P_i$, \cite{Goldstone:1974gf}.  
This is the answer if $P_{i,f}$ are also assumed to satisfy $P_{i,f} \ll M$.  More generally, if $P_{i,f}$ are relativistic, but the transfer is still small, then the leading semiclassical form factor at small momentum transfer is given by the Fourier transform of the boosted soliton profile, $\phi_0(\rho / \sqrt{1-\beta^2})$, where the velocity $\beta$ is determined by the usual relativistic relationship to a momentum $P = \half (P_i + P_f)$: $M_0 \beta/\sqrt{1-\beta^2} = \half (P_i + P_f)$, \cite{Papageorgakis:2014dma}.  
These results, however, can only be trusted to leading order in $k/M$, as they are obtained by dropping the higher-derivative terms in $H_{\rm eff}[P]$.  

In order to probe the soliton-antisoliton pair creation amplitude related to $\llangle \Psi_{P_f} | \widehat{\phi} | \Psi_{P_i} \rrangle$ by crossing symmetry, for example, we must understand the behavior of the form factor for momentum transfers of order the soliton mass, $P_f - P_i \sim O(M)$.  This means that a derivative expansion of $H_{\rm eff}[P]$ is not under control and the all-orders result, obtained from a saddle-point approximation to \eqref{Heffdef} for general $P(t)$, is required.  We will return to this discussion in the final section, where we will obtain an expression for semiclassical soliton form factors in terms of a certain generating functional built from the first two terms in \eqref{Heffexpansion}.

\section{Saddle-point Equation for General $P(t)$}\label{sec:SP}

We will now evaluate, by saddle point approximation, the effective soliton Hamiltonian, \eqref{Heffdef} for general soliton momentum $P(t)$. It is worth recalling how this is done in the small-velocity approximation before tackling the general analysis.  

\noindent{\bf Small Velocity:} If one assumes small soliton velocities,  $\dot{X} = O(g)$ then, since $M_0 = O(g^{-2})$, $\phi_0 = O(g^{-1})$, while $\chi,\varpi = O(1)$, we have the following for the second term in the Hamiltonian \eqref{Ham}:
\begin{equation}\label{smallvelocity}
\frac{ \left( P + \langle \varpi |\chi'\rangle \right)^2}{2 M_0 \left(
    1 + \frac{1}{M_0} \langle \phi_{0}' |\chi'\rangle \right)^2} = \frac{P^2}{2M_0} + O(g)~.
\end{equation}
In this approximation, all field-theory interactions containing fluctuations coming from this term are higher-order in the coupling, compared to the quadratic-order terms coming from the remaining part of the Hamiltonian,\footnote{Note also, however, that the quadratic terms coming from the remaining part of the Hamiltonian are of the \emph{same order} in $g$ as $P^2/2M_0$ in this approximation.  Therefore it is inconsistent to ignore them while keeping the $P^2/2M_0$ term.  They lead, in particular, to the one-loop correction to the soliton mass.} leading to
\begin{equation}\label{Hamsmallg}
H = M_0 +\frac{P^2}{2M_0} + \half \int d\rho \left\{ \varpi^2 + \chi^{\prime 2} + V^{(2)}(\vphi_0) \chi^2 + V_{\delta m^2}(\phi_0) \right\} + O(g)~.
\end{equation}
The $O(1)$ part of the Hamiltonian can be diagonalized by employing the basis of modes \eqref{eq:4}. Inserting these expansions back into \eqref{Heffdef} and working perturbatively in  $g$, one derives explicit Feynman rules for carrying out field-theoretic computations in the soliton sector \cite{Gervais:1975pa}.  These rules include one-point vertices in the fluctuation fields contained in \eqref{smallvelocity}, as well as an infinite series of higher-point vertices. The reason the one-point vertices are present is that $\phi_0(t,x-X(t))$ is not a solution to the classical equations of motion.

\noindent{\bf Constant Velocity:} The authors of \cite{Gervais:1975pa} also demonstrated how one can find the true saddle point of the soliton-sector Lagrangian when the soliton velocity is not small, but constant. In that reference it was shown that
\begin{equation}\label{solconstP}
\chi(\rho) = \phi_0 \left(\frac{\rho}{\sqrt{1-\beta^2}}\right) - \phi_0(\rho)~, \qquad \textrm{with} \quad  P = \frac{M_0 \beta}{\sqrt{1 - \beta^2}}~,
\end{equation}
solves the equations of motion following from \eqref{ssH} when $P$ is time-independent.  The parameter $\beta$ is interpreted as the soliton velocity and has the correct relativistic relationship with $P$.  The solution \eqref{solconstP} is quite nontrivial from the point of view of the equations of motion for $(\chi,\varpi)$ following from \eqref{ssH}, which are nonlocal. It is anticipated from Lorentz invariance of the theory, though, since by \eqref{cxfmphi} it corresponds to the boosted-soliton profile in terms of the original field-theory variable
\begin{equation}
\phi(t,x) = \phi_0 \left( \frac{x -(X_0 + \beta t)}{\sqrt{1-\beta^2}}\right) ~.
\end{equation}
In fact, one could imagine implementing the canonical transformation from the perturbative-sector variables \emph{directly} to variables adapted to the boosted-soliton background from the very beginning; this has indeed been carried out in \cite{Christ:1975wt,Gervais:1975yg,Gervais:1976wr}.

Our goal in this section  will be to generalize the analysis of \cite{Gervais:1975pa} to the case where $P(t)$ is a generic time-dependent function. This will furnish the effective Hamiltonian of the soliton, $H_{\rm eff}[P]$, that is appropriate for processes involving changes in the soliton momentum that are not small, such as soliton form factors with arbitrary momentum transfer. Our approach will follow that of \cite{Gervais:1975pa}. That is, we will first transform to static-soliton variables using \eqref{cxfmphi}, \eqref{cxfmpi} leading to \eqref{Zfi}, \eqref{ssH}, and then find a nontrivial saddle-point solution in those variables. It is natural to ask why we do not directly perform the canonical transformation using a time-dependent soliton background. It turns out that one runs into technical difficulties at the quantum level when attempting to construct a canonical transformation that utilizes a soliton background depending on \emph{both} a time-dependent collective-coordinate position \emph{and} momentum. As briefly discussed under equation~\eqref{Ham}, the resolution of operator-ordering ambiguities in a canonical formalism leads to additional, quantum contributions to the potential. For a generic $(X(t),P(t))$-dependent soliton background these might not be under control.

\subsection{Time-dependent Equations of Motion}

We now begin our discussion in earnest. In this subsection we are simply analyzing classical equations of motion and therefore we work with the theory on $\mathbbm{R}$.  We find it convenient however to use the form of the soliton sector Hamiltonian given in \eqref{HamL}.  This is identical to \eqref{Ham} for the theory on $\mathbbm{R}$, utilizing \eqref{eq:zeromode} and the constraints.  We work with the shifted field\footnote{It might seem that we are in effect undoing the canonical transformation.  This is not the case.  Unlike the original field $\phi$, the field $\varphi$ satisfies a constraint and is independent of the soliton collective coordinate.}
\begin{equation}
\varphi(t,\rho) := \chi(t,\rho) + \phi_0(\rho) ~.
\end{equation}
The transformation $(\chi,\varpi) \mapsto (\varphi,\varpi)$ is canonical, and \eqref{HamL} takes the form
\begin{align}\label{ssH2}
e^{-i \int dt  H_{\rm eff}[P]} =&~ \int [D\varphi D\varpi D\lambda D\nu ]\Psi_{0}[\varphi(t_f,\rho)]^* \Psi_{0}[\varphi(t_i,\rho)] e^{i \int dt (\langle \varpi |\dot{\varphi}\rangle - H_T) } ~, \cr
H_T =&~ \lambda \langle\uppsi_0 |\varphi - \phi_0 \rangle + \nu \langle\uppsi_0 |\varpi \rangle+ \cr
&~ +  \frac{\left( P + \langle\varpi |\varphi' \rangle\right)^2}{2 \langle\uppsi_0 |\varphi'\rangle^2} + \int d\rho \left\{ \half \varpi^2 + \half \varphi^{\prime 2} + V_0(\varphi) + V_{\Delta m^2}(\varphi) \right\} ~. \quad
\end{align}

Treating $P(t)$ as a background variable, the equations of motion following from $H_T$ are\footnote{We do not vary $V_{\Delta m^2}$ when constructing the saddle point solution because terms in $V_{\Delta m^2}$ are suppressed by $O(g^2)$ relative to their counterparts in $V_0$.}
\begin{align}\label{LMsolve1}
  \langle\uppsi_0 |\varphi \rangle =~ \langle \uppsi_0 | \phi_0 \rangle  ~,\qquad \langle\uppsi_0| \varpi\rangle = 0 ~,
\end{align}
and
\begin{align}\label{eomprea}
  \dot{\varphi} =&~ \varpi + \nu \uppsi_0 +  \frac{ P +\langle \varpi |\varphi'\rangle }{\langle\uppsi_0 |\varphi'\rangle ^2} \varphi' ~,\cr
\dot{\varpi} =&~ -\lambda \uppsi_0 + \frac{ P + \langle\varpi |\varphi'\rangle }{\langle\uppsi_0| \varphi' \rangle^2} \varpi' + \varphi'' - V_{0}^{(1)}(\varphi) - \frac{ (P + \langle \varpi |\varphi'\rangle )^2}{\langle\uppsi_0 |\varphi'\rangle^3} \uppsi_0' ~.
\end{align}
At this stage, it is convenient to introduce the ``soliton velocity functional''
\begin{equation}\label{adef}
\beta[\varphi,\varpi;P] := \frac{P + \langle\varpi |\varphi'\rangle}{\langle\uppsi_0 |\varphi'\rangle^2} ~,
\end{equation}
so that the equations of motion \eqref{eomprea} can  be recast into the form
\begin{align}\label{LMsolve2}
  \dot{\varphi} =&~ \varpi + \nu \uppsi_0 + \beta \varphi' ~,\cr
                   \dot{\varpi} =&~ -\lambda \uppsi_0 + \beta\varpi ' + \varphi'' - V_{0}^{(1)}(\varphi) - \uppsi_0' \beta^2 \langle\uppsi_0 |\varphi'\rangle ~. 
\end{align}
The quantity $\beta$ is appropriately named since Hamilton's equation for $X$ in the full theory gives $\dot{X} = \pd H_T/\pd P = \beta$.

Let $(\lambdabar, \nubar,\varphibar,\varpibar)$ denote a solution to these equations and $\abar := \beta[\varphibar,\varpibar;P]$ the velocity functional evaluated on the solution.   The $\nu$ constraint implies that $\dot{\varpibar}$ is orthogonal to $\uppsi_0$, and decay properties of the zero mode imply that $\uppsi_0'$ is orthogonal to $\uppsi_0$ as well.  Thus, the Lagrange multipliers on the solution are determined to be
\begin{equation}\label{backgroundLMs}
\nubar = - \abar\langle \uppsi_0 |\varphibar'\rangle ~, \qquad \lambdabar = \abar\langle \uppsi_0 |\varpibar' \rangle+ \langle\uppsi_0 |(\varphibar'' - V_{0}^{(1)}(\varphibar))\rangle ~.
\end{equation}
By inserting the equation for $\nubar$ back into the $\dot{\varphibar}$ equation and integrating both sides against $\varphibar'$, we can solve for the quantity $\langle \varpibar |\varphibar'\rangle$ and hence determine the velocity $\abar$ purely in terms of $\varphibar$:
\begin{equation}\label{abar}
\abar = \frac{P + \langle\dot{\varphibar}| \varphibar' \rangle}{\langle\varphibar'|\varphibar'\rangle} ~.
\end{equation}

We can now use the $\dot{\varphibar}$ equation to solve for $\varpibar$ in terms of $\varphibar$.  It is convenient to introduce $\PP_\perp$, the linear operator that projects onto functions orthogonal to $\uppsi_0$
\begin{equation}
\PP_\perp(f) := f - \uppsi_0 \langle f |\uppsi_0\rangle~,
\end{equation}
in terms of which the solution for $\varpibar$ is
\begin{equation}\label{backgroundvarpi}
\varpibar = \dot{\varphibar} - \abar \PP_\perp (\varphibar') ~.
\end{equation}
We note that $\PP_\perp(\dot{\varphibar}) = \dot{\varphibar}$, which follows from acting with a time derivative on the $\lambda$ constraint.  Therefore this expression for $\varpibar$ is indeed orthogonal to $\uppsi_0$.

With \eqref{backgroundvarpi} in hand, one can compute $\dot{\varpibar}$, $\varpibar'$, and express the $\dot{\varpibar}$ equation in terms of $\varphibar$ only.  The result can be put in the form
\begin{equation}
\PP_\perp \left( \ddot{\varphibar} - 2\abar \dot{\varphibar}' - \dot{\abar} \varphibar' - (1-\abar^2) \varphibar'' + V_{0}^{(1)}(\varphibar) \right) = 0 ~,
\end{equation}
which implies that
\begin{equation}\label{Cfunctional}
 \ddot{\varphibar} - 2\abar \dot{\varphibar}' - \dot{\abar} \varphibar' - (1-\abar^2) \varphibar'' + V_{0}^{(1)}(\varphibar) = C[\varphibar](t) \uppsi_0(\rho)~
 \end{equation}
for some $\rho$-independent functional $C$ of $\varphibar$.  To find an expression for $C$, multiply both sides of \eqref{Cfunctional} with  $\varphibar'$ and integrate over $\rho$:
\begin{align}
& \frac{d^2}{dt^2} \left( \half \left. \varphibar^2 \right|_{-\infty}^{\infty} \right) - \abar \frac{d}{dt} \langle\varphibar^{\prime}|\varphibar^{\prime}\rangle - \dot{\abar} \langle\varphibar^{\prime}|\varphibar^{\prime}\rangle + \half (1-\abar^2) \left. \varphibar^{\prime 2} \right|_{-\infty}^{\infty} + \left. V_{0}(\varphibar) \right|_{-\infty}^{\infty} = C \langle\uppsi_0| \varphibar'\rangle ~.
\end{align}
We can employ \eqref{abar} to get some cancellations in the second and third term resulting in
\begin{equation}
\half (1-\abar^2) \left. \varphibar^{\prime 2} \right|_{-\infty}^{\infty} + \left. V_{0}(\varphibar) \right|_{-\infty}^{\infty} - \dot{P} - \half \left. \dot{\varphibar}^2 \right|_{-\infty}^{\infty} = C \langle\uppsi_0 |\varphibar'\rangle ~.
\end{equation}
At this point, we impose the usual soliton boundary conditions on $\varphi$: it should approach values in the vacuum $V_0(\varphibar) = V_{\rm min}$ as $\rho \to \pm \infty$ and should have finite energy.  The latter requires that $\varphibar'$ and $\dot{\varphibar}$ should go to zero as $\rho \to \pm \infty$. Armed with this information  we arrive at
\begin{equation}
C = - \frac{\dot{P}}{\langle\uppsi_0 |\varphibar'\rangle} ~,
\end{equation}
through which \eqref{Cfunctional} takes the form of the \emph{forced soliton equation}
\begin{equation}\label{fsea}
(\pd_t - \abar \pd_\rho)^2 \varphibar - \pd_{\rho}^2 \varphibar + V_{0}^{(1)}(\varphibar) = - \frac{\dot{P}}{\langle \uppsi_0| \varphibar'\rangle} \uppsi_0 ~.
\end{equation}
Once a solution to \eqref{fsea} is found, we must still ensure that it satisfies the $\lambda$ constraint
\begin{equation}\label{fseb}
\langle\uppsi_0 |\varphibar\rangle = \langle \uppsi_0 | \phi_0 \rangle = \frac{1}{2\sqrt{M_0}} \left. \phi_{0}^2 \right|_{-\infty}^{\infty} ~,
\end{equation}
where we used \eqref{eq:zeromode} in the second step.

Given a solution $\varphibar$ to \eqref{fsea} and \eqref{fseb}, equations \eqref{backgroundLMs} and \eqref{backgroundvarpi} then determine $\nubar$, $\lambdabar$ and $\varpibar$. The expression for $\lambdabar$ can be further simplified  using \eqref{backgroundvarpi} and \eqref{fsea}, leading to
\begin{equation}\label{backgroundsummary}
\varpibar = \dot{\varphibar} - \abar \PP_\perp (\varphibar') ~, \qquad \nubar = -\abar \langle\uppsi_0| \varphibar'\rangle ~, \qquad \lambdabar = \frac{\dot{P}}{\langle \uppsi_0 |\varphibar'\rangle} - \frac{d}{dt} \left( \abar \langle\uppsi_0| \varphibar'\rangle \right) ~.
\end{equation}

As a simple check of these equations, consider the case of constant $P$. In that example the forcing term on the right-hand side of \eqref{fsea} vanishes, and it is consistent to assume that $\varphibar$ is time independent.  From \eqref{abar} $\abar$ is then constant, and \eqref{fsea} reduces to
\begin{equation}\label{constantPeqn}
(1-\abar^2) \varphibar'' - V_{0}^{(1)}(\varphibar) = 0~, \qquad \textrm{(constant $P$ case)}~.
\end{equation}
After changing variables to $\tilde{\rho} = \rho/\sqrt{1-\abar^2}$ in \eqref{fsea}, one recognizes the standard equation of motion for the soliton, and a solution is
\begin{equation}\label{constantPsol}
\varphibar(\rho) = \phi_0 \left( \frac{\rho- \rho_0}{\sqrt{1-\abar^2}}\right)~, \qquad \textrm{(constant $P$ case)}~.
\end{equation}
The integration constant $\rho_0$ must be chosen so that the constraint \eqref{fseb} is satisfied.  In $\phi^4$ theory, for example, one can take $\rho_0 = 0$.  Remembering from \eqref{massform} that $\langle\phi_{0}'|\phi_{0}'\rangle = M_0$, we deduce $\langle\varphibar'|\varphibar'\rangle = M_0/\sqrt{1-\abar^2}$. Replacing the latter in  \eqref{abar} yields the expected relativistic relationship between momentum and velocity, as in equation~\eqref{solconstP}.  Since the solution is time independent, we see from \eqref{backgroundsummary} that $\lambdabar = 0$.  Meanwhile $\nubar$ and $\varpibar$ can be expressed in terms of $\varphibar'$ and the integral $\langle\uppsi_0| \varphibar'\rangle$.  We are unaware of any simple expression for the value of this integral, which is essentially an overlap between the static zero-mode and the boosted zero-mode.

\subsection{Tree-level $H_{\rm eff}[P]$}

Having established the saddle-point equations for general $P(t)$, and the corresponding background solutions as functionals of $\varphibar$, we next expand in fluctuations by writing
\begin{equation}\label{fluctuations}
(\lambda,\nu,\varphi,\varpi) = (\lambdabar + \delta\lambda, \nubar + \delta\nu, \varphibar + \delta\varphi, \varpibar+ \delta\varpi)~,
\end{equation}
and
\begin{equation}
\int [D\lambda D\nu D\varphi D\varpi] = \int [D \delta\lambda D \delta\nu D \delta\varphi D \delta\varpi ]~.
\end{equation}

Before pressing on, let us first verify that this expansion is under control when $g$ is small, by arguing that all background fields are $O(g^{-1})$.  We begin by noting that \eqref{fsea} is consistent with $\varphibar = O(1/g)$.  Since the soliton mass (\ie\ on-shell Hamiltonian) is $O(1/g^2)$ by the scaling argument above \eqref{eq:rescaledS}, and we are \emph{not} assuming that the velocity is small, then both $P$ and $\dot{P}$ are $O(g^{-2})$.  Then every term in \eqref{fsea} will scale as $O(1/g)$ as long as $\varphibar$ does, implying that $\abar = O(1)$.  It then follows from \eqref{backgroundsummary}, and the fact that $\uppsi_0$ is normalized, that $\varpibar$, $\nubar$, and $\lambdabar$ are also $O(g^{-1})$.  Thus, by treating the fluctuations in \eqref{fluctuations} as $O(1)$ variables, the expansion of the field theory action in fluctuations is an expansion in $g$.

Having established the consistency of the expansion \eqref{fluctuations}, we can now revisit \eqref{ssH2}. The leading-order effective Hamiltonian for the soliton, $H_{\rm eff}^{(-2)}[P] = O(g^{-2})$, is given by
\begin{align}\label{Hgminus2}
H_{\rm eff}^{(-2)}[P] =&~ \left( H_T - \int d\rho V_{\Delta m^2} -\langle{ \varpi| \dot{\varphi}}\rangle \right) \bigg|_{(\lambdabar,\nubar,\varphibar,\varpibar)} \cr
=&~  \frac{ (P + \langle{\varpibar |\varphibar'}\rangle )^2}{2\langle{ \uppsi_0 |\varphibar'}\rangle^ 2} + \int d\rho \left\{ \half \varpibar^2 + \half \varphibar^{\prime 2} + V_0(\varphibar) - \varpi \dot{\varphibar} \right\} ~,
\end{align}
where the Lagrange-multiplier terms in the first line of \eqref{ssH2} vanish since the solution satisfies the constraints.  We have explicitly subtracted out the mass counterterm from $H_T$ in the first equality of \eqref{Hgminus2} since it carries a coefficient that is $O(g^2)$ and therefore only begins contributing to $H_{\rm eff}[P]$ at $O(g^0)$.  Utilizing  \eqref{adef}, \eqref{backgroundvarpi} as well as the fact that $\langle{\uppsi_0 |\dot{\varphibar}}\rangle = 0$, we obtain the following expressions for the various terms in \eqref{Hgminus2}:
\begin{align}
\frac{ (P + \langle{\varpibar |\varphibar'}\rangle )^2}{2\langle{\uppsi_0 |\varphibar'}\rangle^2} =&~ \half \abar^2  \langle{\uppsi_0| \varphibar'}\rangle^2 ~, \cr
\half \langle{\varpibar|\varpibar}\rangle =&~ \half   \langle{\dot{\varphibar} - \abar \PP_\perp(\varphibar')|\dot{\varphibar} - \abar \PP_\perp(\varphibar') }\rangle  \cr
=&~ \half  \langle{\dot{\varphibar}|\dot{\varphibar}}\rangle - 2\abar \langle{\dot{\varphibar}| \varphibar'}\rangle + \abar^2 \langle{\varphibar^{\prime}|\varphibar^{\prime}}\rangle  - \half \abar^2 \langle{ \uppsi_0 |\varphibar'}\rangle^2 ~, \cr
\langle{\varpibar| \dot{\varphibar}}\rangle =&~  \langle{\dot{\varphibar} - \PP_\perp (\varphibar')| \dot{\varphibar}}\rangle = \langle{\dot{\varphibar}|\dot{\varphibar}}\rangle - \abar \langle{\dot{\varphibar}| \varphibar'}\rangle  ~.
\end{align}
Hence the tree-level soliton effective Hamiltonian takes the form 
\begin{equation}\label{HeffTreefinal}
H_{\rm eff}^{(-2)}[P]  = \int d\rho \left\{ \half (1+ \abar^2) \varphibar^{\prime 2} - \half \dot{\varphibar}^2 + V_0(\varphibar) \right\}~,
\end{equation}
where $\abar$ is given by \eqref{abar}, and $\varphibar$ should be viewed as a functional of $P$, defined by the solution to \eqref{fsea} and \eqref{fseb}, in which $P(t)$ appears as a background variable.

Without a time-dependent solution to these equations we cannot be more explicit regarding the form of the tree-level $H_{\rm eff}$, but we can check that our result reduces to the correct expression for the case of constant $P$ where the solution for $\varphibar$ is given by  \eqref{constantPsol}.   This follows from two results.  First, ~\eqref{constantPeqn} implies a virial theorem $\int V_0(\varphibar) = \half \sqrt{1-\abar^2} \, M_0$.  Second, 
 $\langle{\varphibar^{\prime }|\varphibar^{\prime }}\rangle = M_0/\sqrt{1-\abar^2}$, and putting these observations together we obtain
\begin{align}\label{Heffm2result}
 H_{\rm eff}^{(-2)}[P] 
 =&~ \half(1 + \abar^2)\frac{M_0}{\sqrt{1-\abar^2}} + \half \sqrt{1-\abar^2} M_0 = \frac{M_0}{\sqrt{1-\abar^2}}  \cr
=&~ \sqrt{P^2 + M_{0}^2} ~, \qquad \textrm{(constant $P$ case)}~, 
\end{align}
as expected.

\subsection{The Semiclassical Correction}

Our next task is to expand \eqref{ssH2} in the fluctuations introduced in \eqref{fluctuations}.  The linear terms vanish\footnote{$V_{\Delta m^2}$ was not included in the extremization and therefore may contain terms linear in the fluctuations.  Indeed, the presence of such terms is necessary to cancel tadpoles generated by cubic interactions in $\delta\chi$, \cite{Gervais:1975pa}.  Such terms will be suppressed in the coupling expansion and do not affect our one-loop analysis.} 
by virtue of $(\lambdabar,\nubar,\varphibar,\varpibar)$ extremizing the action $\int dt ( \langle{\varpi |\dot{\varphi} }\rangle- H_T)$. We therefore examine the quadratic-order terms, setting the stage for the one-loop computation in the following section.

Using \eqref{adef}, and working still on all of $\mathbbm{R}$, the quantity to be expanded can be written as
\begin{align}
\langle{\varpi |\dot{\varphi}}\rangle - H_T  =&~   \langle{\varpi |\dot{\varphi}}\rangle -  \lambda \left( \langle{\uppsi_0 |\varphi }\rangle- \frac{1}{2\sqrt{M_0}} \phi_{0}^2 \bigg|_{-\infty}^{\infty} \right) - \nu\langle{\uppsi_0| \varpi}\rangle + \cr
&~ -\half \beta^2 \langle{\uppsi_0 |\varphi'}\rangle^2 -  \int d\rho \left\{ \half \varpi^2 + \half \varphi^{\prime 2} + V_0(\varphi) + V_{\delta m^2}(\varphibar) \right\} + O(g)~. \quad 
\end{align}
Therefore the terms that are second order in fluctuations are
\begin{align}\label{2ndorder1}
\left(\langle{ \varpi |\dot{\varphi}}\rangle - H_T \right) \bigg|_{O(\delta^2)} =&~  \langle{\delta\varpi| \dot{\delta\varphi}}\rangle - \delta\lambda \langle{\uppsi_0 |\delta\varphi}\rangle - \delta\nu \langle{\uppsi_0 |\delta\varpi}\rangle + \cr
&~ - \frac{1}{4} (\delta^2 \beta^2 |) \langle{\uppsi_0 |\varphibar'}\rangle^2 + 2\abar \delta \beta | \langle{\uppsi_0| \varphibar' }\rangle \langle{\uppsi_0'| \delta\varphi}\rangle - \frac{\abar^2}{2} \langle{\uppsi_{0}' |\delta\varphi}\rangle^2  + \cr
  &~  - \half \langle{\delta\varpi|\delta\varpi}\rangle - \half \langle \delta\varphi | (- \pd_{\rho}^2 + V_{0}^{(2)}(\varphibar) ) \delta\varphi \rangle + O(g)~,
\end{align}
where the vertical bar appearing on the right side represents evaluation on the background solution: $| \equiv |_{(\lambdabar,\nubar,\varphibar,\varpibar)}$.  Since
\begin{equation}
\delta^{2} \beta^2 = 2 \beta \delta^{2} \beta + 2 (\delta \beta)^2~,
\end{equation}
the greatest challenge in this computation lies in obtaining the expansion of $\beta$ to quadratic order in fluctuations. Equation~\eqref{adef} yields
\begin{align}
\delta \beta =&~ \frac{\langle \delta\varpi | \varphi' \rangle + \langle \varpi | \delta\varphi' \rangle}{\langle \uppsi_0 |\varphi' \rangle^2} + \frac{ 2 (P + \langle \varpi | \varphi' \rangle )}{\langle \uppsi_0 |\varphi' \rangle^3} \langle \uppsi_{0}' | \delta\varphi \rangle \cr
\delta^{2} \beta =&~ \frac{2 \langle \delta\varpi | \delta \varphi' \rangle}{\langle \uppsi_0 | \varphi' \rangle^2} + \frac{4 (\langle \delta\varpi | \varphi' \rangle + \langle \varpi | \delta\varphi' \rangle )}{\langle \uppsi_0 | \varphi' \rangle^3} \langle \uppsi_{0}' |\delta\varphi \rangle + \frac{6 (P + \langle \varpi | \varphi' \rangle)}{\langle \uppsi_0 | \varphi' \rangle^4} \langle \uppsi_{0}' |\delta\varphi \rangle^2 ~,
\end{align}
and evaluating the above on the background gives
\begin{align}
\delta \beta | =&~ \frac{ \langle \delta\varpi | \varphibar' \rangle + \langle \varpibar | \delta\varphi' \rangle}{\langle \uppsi_0 |\varphibar' \rangle^2} + \frac{2 \abar \langle \uppsi_{0}' |\delta\varphi \rangle}{  \langle \uppsi_0 |\varphibar' \rangle } ~,  \cr
\delta^2 \beta | =&~ \frac{2 \langle \delta\varpi |\delta\varphi' \rangle }{\langle \uppsi_0 |\varphibar' \rangle^2} + \frac{4 ( \langle \delta\varpi | \varphibar' \rangle + \langle \varpibar | \delta\varphi' \rangle) }{\langle \uppsi_0 |\varphibar' \rangle^3} \langle \uppsi_{0}' |\delta\varphi \rangle + \frac{6\abar  \langle \uppsi_{0}' |\delta\varphi \rangle^2 }{  \langle \uppsi_0 |\varphibar' \rangle^2} ~.
\end{align}

With these expressions in hand, the middle line of \eqref{2ndorder1} can be put in the form
\begin{align}\label{amess}
& - \half \left( \abar \delta^2 \beta | + (\delta \beta | )^2 \right) \langle \uppsi_{0} |\varphibar' \rangle^2 + 2\abar (\delta \beta | ) \langle \uppsi_0 | \varphibar' \rangle \langle \uppsi_{0}' |\delta\varphi \rangle - \frac{\abar^2}{2} \langle \uppsi_{0}' | \delta\varphi \rangle^2 = \cr
& \qquad = - \half \bigg\{  \frac{ (\langle \delta\varpi | \varphibar' \rangle + \langle \varpibar |\delta\varphi' \rangle )^2}{\langle \uppsi_0 |\varphibar' \rangle^2} + 2\abar \langle \delta\varpi | \delta\varphi' \rangle + \frac{4\abar (\langle \delta\varpi | \varphibar' \rangle + \langle \varpibar |\delta\varphi' \rangle )}{\langle \uppsi_0 |\varphibar' \rangle} \langle \uppsi_{0}' |\delta\varphi \rangle + \cr
&~ \qquad \qquad \qquad + 3 \abar^2 \langle \uppsi_0' |\delta\varphi \rangle^2 \bigg\} ~.
\end{align}
Inserting this back into \eqref{2ndorder1} and collecting terms, we find that 
\begin{align}\label{quadflucfinal}
  \left(\langle{ \varpi| \dot{\varphi}}\rangle - H_T \right) \bigg|_{\delta^{(2)}} =&~ \langle \delta\varpi | \dot{\delta\varphi} \rangle - \delta\lambda \langle \uppsi_0 | \delta\varphi \rangle - \delta\nu \langle \uppsi_0 |\delta\varpi \rangle + \cr
&~ - \half \langle \delta\varpi | \MM \delta\varpi \rangle - \langle \delta\varpi | \BB \delta\varphi \rangle - \half \langle \delta\varphi | \KK \delta\varphi \rangle + O(g)~,
\end{align}
where the linear operators $\MM$, $\BB$, and $\KK$ are given by
\begin{align}\label{opersabove}
\MM :=&~ \mathbbm{1} + \frac{ |\varphibar' \rangle \langle \varphibar' |}{\langle \uppsi_0 | \varphibar' \rangle^2} ~, \cr 
\BB :=&~   \abar \pd_\rho + \frac{ |\varphibar' \rangle \langle \abar \varphibar'' - \dot{\varphibar}' | }{\langle \uppsi_0 |\varphibar' \rangle^2} + \frac{\abar |\varphibar' \rangle \langle \uppsi_0' | }{\langle \uppsi_0 |\varphibar' \rangle }~, \cr
\KK :=&~ - \pd_{\rho}^2 + V_{0}^{(2)}(\varphibar) + \frac{ |\abar \varphibar'' - \dot{\varphibar}' \rangle \langle \abar \varphibar'' - \dot{\varphibar}' | }{ \langle \uppsi_0 |\varphibar' \rangle^2} + \frac{\abar \left( |\abar \varphibar'' - \dot{\varphibar}' \rangle \langle \uppsi_0' | + |\uppsi_0' \rangle \langle \abar \varphibar'' - \dot{\varphibar}' | \right) }{\langle \uppsi_0 |\varphibar' \rangle} ~. \cr
\end{align}
Here we used \eqref{backgroundsummary} to set $\varpibar' = \dot{\varphibar}' - \abar \varphibar'' + \abar \uppsi_{0}' \langle \uppsi_0 | \varphibar' \rangle$ in several places.  

In order to arrive at \eqref{opersabove} we integrated by parts on various terms in \eqref{amess} that involve $\varphibar'$ integrated against the fluctuations.  Thus for the theory in the box, we have neglected terms in \eqref{opersabove} of order $\varphibar'(t,\pm L/2)$.  We expect, for all $t$, this quantity to be exponentially suppressed in $m L$ when $L$ is large, and therefore we will not concern ourselves with these terms.
  
We can now use  \eqref{quadflucfinal} to determine the $O(1)$ semiclassical correction to $H_{\rm eff}[P]$, \eqref{ssH2}, which we denote by $H_{\rm eff}^{(0)}[P]$.  We find
\begin{align}
  e^{-i \int_{t_i}^{t_f} dt H_{\rm eff}^{(0)}[P]} =& \int [D\delta\lambda D\delta\nu D\delta\varphi D\delta\varpi] \Psi_{0}[\delta\varphi(t_f,\rho)]^*\Psi_{0}[\delta\varphi(t_i,\rho)]\times \cr
                                         & \times \exp \bigg\{ i \int_{t_i}^{t_f} dt \bigg(  \langle \delta\varpi | \dot{\delta\varphi} \rangle - \delta\lambda \langle \uppsi_0 | \delta\varphi \rangle - \delta\nu \langle \uppsi_0 |\delta\varpi \rangle + \cr
&~ \qquad \quad   - \half \langle \delta\varpi | \MM \delta\varpi \rangle - \langle \delta\varpi | \BB \delta\varphi \rangle - \half \langle \delta\varphi | \KK \delta\varphi \rangle - V_{\delta m^2}(\varphibar) \bigg) \bigg\} ~. \qquad
\end{align}
The form of the Lagrange-multiplier terms in this expression strongly suggests that we should expand the fluctuation fields in the orthonormal basis $\{\uppsi_n \}$:
\begin{equation}\label{accelmodeexp}
\delta\varphi(t,\rho) = \sum_n q_n(t) \uppsi_n(\rho)~, \qquad \delta\varpi(t,\rho) = \sum_n p_n(t) \uppsi_n(\rho)~.
\end{equation}
We recall from \eqref{eq:3} that these modes diagonalize the fluctuation operator around the \emph{static} soliton.  While they certainly do not diagonalize the quadratic Hamiltonian under current investigation (unless $P = 0$), they \emph{do} allow us to cleanly dispose of the constraints: integrating over $\delta\lambda$ and $\delta\nu$ produces the product of Dirac delta functionals $\delta(q_0(t)) \delta(p_0(t))$.  Since
\begin{equation}
[D\delta\varphi D \delta\varpi] = \prod_{n=0} [Dq_n Dp_n] ~,
\end{equation}
one can then soak up the delta functionals by integrating over $q_0$ and $p_0$.  This effectively removes $q_0,p_0$ from the mode expansions \eqref{accelmodeexp}, so that the problem only depends on the restriction of the operators $\MM$, $\BB$, $\KK$ to the orthogonal complement of $\mathrm{Span}\{ |\uppsi_0\rangle \}$.

It will be useful to put the remaining integrations into a standard form.  We define column vectors $\bq, \bp$ with components $q_n,p_n$ and matrices $\cbM,\cbB,\cbK$ with real components\footnote{In the following, boldfaced quantities will always refer to the matrix representation of that quantity with respect to the basis $\{ |\uppsi_n \rangle \}_{n \neq 0}$.}
\begin{equation}
\MM_{mn} := \langle \uppsi_m | \MM \uppsi_n \rangle~, \qquad \BB_{mn} := \langle \uppsi_m | \BB \uppsi_n \rangle~, \qquad \KK_{mn} := \langle \uppsi_m |\KK \uppsi_n \rangle~,
\end{equation}
for $m,n = 1,\ldots, N$, where $N$ is the total number of modes we consider.\footnote{The relationship between $N$ and the UV cut-off employed for mass renormalization in the perturbative sector will be discussed in subsection \ref{sec:oneloopcp} below.}  We also collect these matrices into a $2N \times 2N$ symmetric real block matrix
\begin{equation}
\label{eq:FluctuationHamiltonian}
\cbH := \left( \begin{array}{c c} \cbM & \cbB \\ \cbB^{T} & \cbK \end{array} \right) ~,
\end{equation}
and we write $[D\bq D\bp] = \prod_{n=1}^{N} [Dq_n Dp_n]$ for the phase space measure.  Then
\begin{align}\label{HeffOneloop}
  \exp\left\{-i \int_{t_i}^{t_f} d t H_{\rm eff}^{(0)}[P] \right\} =& ~ \exp \left\{- i \int_{t_i}^{t_f} d t V_{\delta m^2}(\varphibar)\right\} \, \cI[P;t_f, t_i]~, 
 \end{align}
 where
\begin{align}\label{IPI}
\cI[P;t_f,t_i] :=& \int [D\bq D\bp] \Psi_{0}(\bq_f)^*\Psi_{0}(\bq_i)  \exp \left\{ i \int_{t_i}^{t_f} dt \left( \bp^T \dot{\bq} - \half \begin{pmatrix} \bp^T & \bq^T \end{pmatrix} \cbH \begin{pmatrix} \bp \\ \bq \end{pmatrix} \right) \right\} ~,
\end{align}
with $\bq_{i,f} \equiv \bq(t_{i,f})$.  The $P$ dependence of $\II$ comes through the quadratic form $\cbH$, which depends on $P$ through the background solution $\varphibar$.  The next section is concerned with analyzing this quadratic path integral in detail.  

Recall that $V_{\delta m^2}(\varphibar)$ is the one-loop mass counterterm from the perturbative sector, evaluated on the soliton background.  In the case of constant $P$ it was shown in \cite{Jain:1990dq}, generalizing the classic computation of \cite{Dashen:1974cj}, how this counterterm renders the one-loop correction to the relativistic soliton energy finite in $\phi^4$ theory.  We will review, clarify, and expand on this computation at the end of the next section.

\section{One-loop Correction}\label{sec:SHO}

Starting with the field theory path integral of the previous section, we obtained a finite-dimensional
quantum mechanics by working in a system of fixed spatial size $L$ and imposing a cut-off on the mode number.  Since the resulting action is quadratic in coordinates and momenta, the path integral can be evaluated explicitly.  However, the Hamiltonian we have obtained is slightly unusual: it has a term linear in coordinates and momenta.  Fortunately, much of the technology, even if unfamiliar, was developed long ago.  We will collect and review the pertinent results and then apply them to the quantum mechanics of the discretized fluctuation path integral of our field theory.

The quantum mechanical path integral we are after, \eqref{IPI}, can be expressed in terms of the standard transition amplitude, or propagator,
\begin{align}
Z(\bq_f,\bq_i;t_f,t_i) :=&~ \int_{\bq(t_i)=\bq_i}^{\bq(t_f)=\bq_f} [D\bq D\bp] \exp \left\{ i \int_{t_i}^{t_f} dt \left( \bp^T \dot{\bq} - \half \begin{pmatrix} \bp^T & \bq^T \end{pmatrix} \cbH \begin{pmatrix} \bp \\ \bq \end{pmatrix} \right) \right\}~,
\end{align}
via
\begin{equation}\label{IPI2}
\cI[P;t_f,t_i] = \int d^N \bq_f d^N \bq_i  \Psi_{0}(\bq_f)^* Z(\bq_f,\bq_i;t_f,t_i) \Psi_{0}(\bq_i) ~.
\end{equation}
Hence there are two pieces to the calculation: the propagator and the initial and final state wavefunctions.  In the next three subsections we will reduce \eqref{IPI2} to a finite-dimensional determinant.  The result is quite general and does not utilize the detailed form of $\cbH$.  In subsection \ref{sec:spectrumcp} we diagonalize our explicit $\cbH$ in the case of constant $P$, and in subsection \ref{sec:oneloopcp} we put all these results together to obtain the expected form of the one-loop correction to the relativistic soliton energy.
 
\subsection{The Propagator}

Consider the quantum mechanics of the $N$ degrees of freedom with conjugate momentum and position operators $\bph$ and $\bqh$ and quadratic Hamiltonian\footnote{We will use hats to distinguish quantum mechanical operators from the corresponding classical quantities.}
\begin{align}
\Hh & = \frac{1}{2} \begin{pmatrix} \bph^T & \bqh^T \end{pmatrix} \cbH \begin{pmatrix} \bph \\ \bqh \end{pmatrix} ~,
\end{align}
and corresponding time evolution operator
\begin{align}
\Uh(t_f,t_i) &=  \boldsymbol{\text{T}} \left\{ \exp \left[-i\int_{t_i}^{t_f} d t\, \Hh(t) \right] \right\}~,
\end{align}
with $\boldsymbol{\text{T}}$ the time-ordering operator.  

We will need to recall some details of the path integral computation of the propagator for this system, \ie\,the matrix element
\begin{align}
Z(\bq_f,\bq_i;t_f,t_i)  = \la \bq_f | \Uh(t_f,t_i) |  \bq_i \ra   = \int_{\bq(t_i)=\bq_i}^{\bq(t_f) = \bq_f} [D\bq D \bp] \, e^{i S}~.
\end{align}
We will implicitly work with the mid-point prescription for the phase-space path integral, where the time interval $t_i \le t \le t_f = t_i + T$ is divided into $T/\epsilon$ segments of step size $\epsilon$, with the coordinate variables defined on the end-points of the intervals, while the momenta are defined at the midpoints of the intervals.  We refer the reader to standard references such as the modern and thorough text~\cite{Grosche:1998yu} for results, history and references on quantum--mechanical path integrals.  In particular, for any action $S$ quadratic in $\bq$ and $\bp$, possibly with time-dependent coefficients, the path integral can be evaluated in closed form as
\begin{align}
\label{eq:generalquadraticPI}
Z(\bq_f,\bq_i;t_f,t_i) = \left(\frac{1}{2\pi i}\right)^{N/2} \sqrt{ \det_{m,n} \left( -\frac{\p^2 S}{\p q_{f}^{m} \p q_{i}^{n}} \right)}\, e^{i S}~,
\end{align}
where $S = S(\bq_f,\bq_i;t_f,t_i)$ is the action evaluated on the classical solution to the boundary value problem $\bq(t_i) = \bq_i$ and $\bq(t_f) = \bq_f$.  (See subsection 6.2 of \cite{Grosche:1998yu} for an extensive list of references.)

To evaluate the classical action in a useful form, we recall some key points from classical mechanics with quadratic Hamiltonians.  Letting
\begin{align}
\cbJ & = \begin{pmatrix}
0 & -\iden \\ \iden & ~0
\end{pmatrix}~,
\end{align}
denote the symplectic structure, the equations of motion for the quadratic Hamiltonian,
\begin{align}
H & = \frac{1}{2} \begin{pmatrix} \bp^T & \bq^T \end{pmatrix} \cbH \begin{pmatrix} \bp \\ \bq \end{pmatrix} ~,
\end{align}
are
\begin{align}\label{classicaleoms}
\begin{pmatrix}
\dot \bp \\ \dot \bq 
\end{pmatrix}
= \cbJ \cbH \begin{pmatrix}
\bp \\ \bq
\end{pmatrix} ~.
\end{align}
Given some initial data $\bq_i = \bq(t_i)$ and $\bp_i = \bp(t_i)$, the solution takes the form
\begin{align}
\label{eq:generaltimedependentsolution}
\begin{pmatrix}
\bp_f \\ \bq_f 
\end{pmatrix} 
= \boldsymbol{\text{T}} \left\{ \exp \left[\int_{t_i}^{t_f} d t\, \cbJ \cbH(t) \right] \right\}
\begin{pmatrix}
\bp_i \\ \bq_i
\end{pmatrix}~.
\end{align}

We note that all the usual Hamilton-Jacobi manipulations simplify for quadratic $H$.  Quite generally, we have
\begin{align}
\bp^T \dot\bq = \frac{1}{2} \frac{d}{dt} (\bp^T \bq) - \frac{1}{2} \begin{pmatrix}
\bp^T & \bq^T
\end{pmatrix}  \cbJ \begin{pmatrix}
\dot \bp \\ \dot \bq
\end{pmatrix}  = \frac{1}{2} \frac{d}{dt} (\bp^T \bq) + \frac{1}{2}\sum_{j=1}^N \left( p^j \frac{\p H}{\p p^j} + q^a \frac{\p H}{\p q^j} \right)~.
\end{align}
We used the equations of motion in the second equality.  Thus,
\begin{align}
\bp^T \dot\bq - H & =  \frac{1}{2} \frac{d}{dt} (\bp^T \bq) + \frac{1}{2} \sum_{j=1}^N\left( p^j \frac{\p H}{\p p^j} + q^j \frac{\p H}{\p q^j} \right) - H~,
\end{align}
and the last two terms cancel when $H$ is homogeneous of degree $2$ in the $\bp$ and $\bq$.  The classical action is therefore simply
\begin{align}
S = \int_{t_i}^{t_f} d t \left( \bp^T \dot\bq - H \right)  =  \frac{1}{2} \left(\bp_f^T \bq_f - \bp_i^T \bq_i\right)~. 
\end{align}

To apply this to the path integral in \eqref{eq:generalquadraticPI}, we need to use the general solution \eqref{eq:generaltimedependentsolution} to express the $\bp_f$ and $\bp_i$ variables in terms of the $\bq_f$ and $\bq_i$.  Writing the classical evolution operator in $N\times N$ block form as
\begin{align}\label{classicalevolution}
\boldsymbol{\text{T}} \left\{ \exp \left[\int_{t_i}^{t_f} d t\, \cbJ \cbH(t) \right] \right\} = \begin{pmatrix} \bM_1 & \bM_2 \\ \bM_3 & \bM_4 \end{pmatrix}~,
\end{align}
we have
\begin{align}
\bp_f & = \bM_1 \bp_i + \bM_2 \bq_i~,&
\bq_f & = \bM_3 \bp_i + \bM_4 \bq_i~.
\end{align}
Assuming that $\bM_3$ is invertible,\footnote{We expect this to hold for generic $T$, and will at any rate see that the expressions we obtain for our path integral will not be affected by this assumption: all factors of $\bM_3^{-1}$ will cancel in the final expressions.} we obtain
\begin{align}
\begin{pmatrix}
\bp_f \\ \bp_i 
\end{pmatrix}
=
\begin{pmatrix}
\bM_1 \bM_3^{-1} ~&~ \bM_2 - \bM_1 \bM_3^{-1} \bM_4 \\
\bM_3^{-1} ~&~ -\bM_3^{-1} \bM_4
\end{pmatrix}
\begin{pmatrix}
\bq_f \\ \bq_i 
\end{pmatrix}~.
\end{align}
Plugging this into the action, we obtain
\begin{align}
\label{eq:evaluatedaction}
S(\bq_f,\bq_i;t_f,t_i) & = \frac{1}{2} \begin{pmatrix} \bq_f^T & \bq_i^T \end{pmatrix} \cbS
\begin{pmatrix} \bq_f \\ \bq_i \end{pmatrix}~,& \textrm{where} \quad
\cbS & = \begin{pmatrix}
\cbS_{ff} & \cbS_{fi} \\ \cbS_{fi}^T & \cbS_{ii}
\end{pmatrix}~,
\end{align}
and the blocks of the $2N\times2N$ symmetric matrix $\cbS$ are
\begin{align}
\cbS_{ff} & = \ff{1}{2}( \bM_1 \bM_3^{-1} + (\bM_3^{-1})^T \bM_1^T)~, \nonumber\\
\cbS_{fi} & = \ff{1}{2}( \bM_2 - \bM_1 \bM_3^{-1} \bM_4 - (\bM_3^{-1})^T)~, \nonumber\\
\cbS_{ii} & = \ff{1}{2} (\bM_3^{-1} \bM_4 + \bM_4^T (\bM_3^{-1})^T)~.
\end{align}
Then the Morette--Van Hove determinant is read off as
\begin{align}
\sqrt{ \det_{m,n} \left( -\frac{\p^2 S}{\p q_f^m \p q_i^n} \right)} = \sqrt{\det(- \cbS_{fi}) }~.
\end{align}

The classical time evolution operator is a symplectic transformation on the phase space (this holds even when the Hamiltonian is time-dependent), so that the matrices $\bM_1$, $\bM_2$, $\bM_3$, and $\bM_4$ satisfy the $\Sp(2N,\R)$ identities that the products
$\bM_1^T \bM_3$,  $\bM_1 \bM_2^T$, $\bM_2^T \bM_4$, and $\bM_4 \bM_3^T$ are symmetric, and
\begin{align}
\bM_4^T \bM_1 - \bM_2^T \bM_3 & = \iden~, &
\bM_4 \bM_1^T - \bM_3 \bM_2^T & = \iden~.
\end{align}
This leads to
\begin{align}
\cbS_{ii} & = \bM_{3}^{-1} \bM_4 = \bM_4^T (\bM_{3}^{-1})^T~, &
\cbS_{ff} & = \bM_1 \bM_3^{-1} = (\bM_3^{-1})^{T} \bM_1^T~,
\end{align}
and we also obtain
\begin{align}
\cbS_{fi} & =  \ff{1}{2} (\bM_2 - \bM_1 \bM_3^{-1} \bM_4 - (\bM_3^{-1})^{T}) = 
\ff{1}{2} (\bM_2 \bM_3^T - \bM_1 \bM_3^{-1} \bM_4 \bM_3^{T} -\iden) (\bM_3^{-1})^{T}  \nonumber\\
& = \ff{1}{2} (\bM_2 \bM_3^T - \bM_1 \bM_4^T -\iden ) (\bM_3^{-1})^{T} \nonumber\\
& = -(\bM_3^{-1})^{T}~.
\end{align}
Putting these results together, we obtain
\begin{align}
\label{eq:propagatorfinalform}
Z(\bq_f,\bq_i;t_f,t_i) & = \left(\frac{1}{2\pi i}\right)^{N/2}\frac{1}{\sqrt{\det \bM_3}} e^{i S(\bq_f,\bq_i;t_f,t_i) }~,
\end{align}
with the classical action given in \eqref{eq:evaluatedaction}.

\subsection{The Fluctuation Path Integral for Constant $P$}

The next step is to determine the ground state wavefunctions, $\Psi_{0}(\bq_{i,f})$, and use them with \eqref{eq:propagatorfinalform} to compute \eqref{IPI2}.  It is useful to do this for the case of constant $P$ first, before tackling the general case in the next subsection.

We are assisted by one more piece of machinery from the classical mechanics of quadratic Hamiltonians.  (See, for instance, Appendix 6 of~\cite{Arnold:mech} for a discussion of normal forms of quadratic Hamiltonians.)  Assuming that $\cbH$ is positive definite,\footnote{The explicit diagonalization in subsection \ref{sec:spectrumcp} will justify this assumption.} it was shown by Williamson in~\cite{MR1507138} that there exists a symplectic transformation $\cbC \in \Sp(2N,\R)$ such that
\begin{align}\label{HHWdiag}
\cbC^T \cbH \cbC = \cbN = \begin{pmatrix} \boldsymbol{\nu} & 0 \\ 0 & \boldsymbol{\nu} \end{pmatrix}~,
\end{align}
where $\boldsymbol{\nu}$ is a diagonal matrix, $\boldsymbol{\nu} = \diag(\nu_1,\nu_2,\ldots, \nu_n)$, with $\nu_a >0$.\footnote{ An elegant proof of this result is given in~\cite{Simon:1998abc}:  the matrix $\cbM = \cbH^{-1/2} \cbJ \cbH^{-1/2}$ is invertible and anti-symmetric, and therefore there exists a transformation $\cbR \in \SO(2n,\R)$ and a positive diagonal matrix $\cbN$, such that $\cbR^T \cbM \cbR =  \cbN^{-1} \cbJ$; $\cbC = \cbH^{-1/2} \cbR \cbN^{1/2} $ is the desired symplectic transformation.}  In general both $\boldsymbol{\nu}$ and $\cbC$ will be functions of time, and in this case it is still not straightforward to evaluate the evolution operator, \eqref{classicalevolution}, in a more explicit form.  However, when $\cbH$ is time-independent, as is the case for constant $P$, we can use the symplectic transformation to recast the initial value problem in terms of the new variables $\bp'$, $\bq'$ defined by
\begin{align}
\begin{pmatrix}
\bp' \\ \bq'
\end{pmatrix}
= \cbC^{-1}
\begin{pmatrix}
\bp \\ \bq
\end{pmatrix}~.
\end{align}
Given initial values $\bp'_i = \bp'(t_i)$ and $\bq'_i = \bq'(t_i)$, we find at time $t_f$
\begin{align}
\label{eq:transformedsolution}
\begin{pmatrix}
\bp'(t_f) \\ \bq'(t_f)
\end{pmatrix} =
\underbrace{\begin{pmatrix}
\cos \bnu T & -\sin \bnu T \\ \sin \bnu T & \cos\bnu T
\end{pmatrix}}_{= \cbR}
\begin{pmatrix}
\bp'_i\\ \bq'_i
\end{pmatrix}~. 
\end{align}
Finding either the explicit form of $\cbC$ or of $\bnu$ is not straightforward, even if one has a complete solution to the problem, \eqref{eq:FluctuationHamiltonian}, with $\cbB = 0$,\footnote{It is easy to see that $\det \bnu = \sqrt{\det \cbH}$, but the individual eigenvalues are not related to those of $\cbH$ in any transparent fashion.} but these results will allow us to evaluate \eqref{IPI2} in a simple and useful closed form. 

The Williamson transformation gives us a simple way to describe the soliton ground state $|\Psi_{0}\ra$.   The Williamson transformation
\begin{align}
\cbC = \begin{pmatrix} \bC_1 & \bC_2 \\ \bC_3 & \bC_4\end{pmatrix}
\end{align}
relates momentum and position operators in the quantum mechanics to canonically conjugate operators $\bph'$ and $\bqh'$ via
\begin{align}
\label{eq:newconjugateoperators}
\bph & = \bC_1\bph' + \bC_2 \bqh'~, &
\bqh & = \bC_3 \bph' +\bC_4 \bqh'~,
\end{align}
with the Hamiltonian operator given by
\begin{align}
\label{eq:newhamiltonian}
\Hh & = \frac{1}{2} \begin{pmatrix} \bph^T & \bqh^T \end{pmatrix} \cbH \begin{pmatrix} \bph \\ \bqh \end{pmatrix} =\frac{1}{2} \sum_{a=1}^N \nu_a \left(  \ph'^a \ph'^a + \qh'^a \qh'^a \right)~. 
\end{align}
The creation and annihilation operators
\begin{align}
\label{eq:creationannihilation1}
\bah & = \frac{1}{\sqrt{2}} \left( \bph'- i\bqh'\right) ~,&
\bah^\dag & = \frac{1}{\sqrt{2}} \left( \bph'+i \bqh'\right)~,
\end{align}
satisfy the usual relations
\begin{align}
\label{eq:creationannihilation2}
\CO{\ah^a}{\ah^b} & = 0 ~,&
\CO{\ah^a}{\ah^{\dag b}} & = \delta^{ab}~,&
\Hh & = \sum_{a=1}^N \nu_a \left( \ah^{\dag a} \ah^a+ \frac{1}{2} \right)~.
\end{align}
The ground state $|\Psi_0\ra$ is then defined to be the normalized state annihilated by all $\ah^a$.

With that preparation, we interpret the fluctuation path integral, \eqref{IPI2}, as a matrix element of the time evolution operator in the quantum mechanics and conclude
\begin{align}
\label{eq:constantPresultI}
\cI (T) =&~ \int d^N \bq_f d^N \bq_i  \Psi_{0}(\bq_f)^*\Psi_{0}(\bq_i) Z(\bq_f,\bq_i;t_f,t_i) \cr
=&~ \int d^N \bq_f d^N \bq_i \llangle \Psi_{0} | \bq_f \rrangle \llangle \bq_f | \Uh(t_f,t_i) | \bq_i \rrangle \llangle \bq_i | \Psi_{0} \rrangle    \cr
=&~  \la \Psi_{0} | \Uh(T) |\Psi_{0} \ra = \la \Psi_0 | e^{-i \Hh (t_f -t_i)} |\Psi_0 \ra = \exp\left\{ -\frac{i}{2} T \sum_{a=1}^N \nu_a \right\}~.
\end{align}
The eigenvalues $\nu_a$ will be determined in subsection \ref{sec:spectrumcp}. 

\subsection{The Fluctuation Path Integral for General $P(t)$}\label{sec:PIgen}

More generally, when $\dot P \neq 0$, we face a more complicated evolution problem, but we can nevertheless reduce \eqref{IPI2} by applying the ideas developed above.

First, we observe that the Williamson transformation that allowed us to recast the Hamiltonian operator in diagonal form can still be performed in the time dependent case:  the results in~(\ref{eq:newconjugateoperators}), (\ref{eq:newhamiltonian}), (\ref{eq:creationannihilation1}), and (\ref{eq:creationannihilation2}) continue to hold, with the crucial difference that now $\cbC(t)$, $\bph'(t)$, $\bqh'(t)$, as well as the creation and annihilation operators, are time-dependent,\footnote{In the general case it is difficult to assess whether $\cbH(t)$ is positive-definite at generic $t$.  However the final result we obtain will only depend on the Williamson transformation at initial and final times.  Restricting to $P(t)$ that is constant at early and late times, the background $\varphibar$ will be constant at early times, and hence the results of the next subsection will demonstrate that $\cbH(t_i)$ is positive definite.  At late times, the background solution will approach a superposition of the kink and a spectrum of traveling waves \cite{Faddeev:1977rm}.  We also expect positive definiteness in this case.} as are the corresponding eigenkets, \eg\,the ``position'' eigenkets $|\xi'_{\bq'}(t)\ra$ that satisfy
\begin{align}
\bqh' |\xi'_{\bq'}(t)\ra = \bq' |\xi'_{\bq'}(t)\ra~.
\end{align}
As explained in~\cite{Arvind:1995ab}, since the symplectic transformation~(\ref{eq:newconjugateoperators}) preserves the commutation relations of the $\bqh'$ and $\bph'$, there must be a unitary transformation $\bVh$ that relates the operators:
\begin{align}
\bph' & = \bVh^\dag \bph \bVh~,&
\bqh' & = \bVh^\dag \bqh \bVh~,
\end{align}
so that the time-dependent eigenkets $|\xi'_{\bq'}(t)\ra$ can be expressed in terms of the eigenkets of $\bqh$, which we denote as $|\bq\ra$:
\begin{align}
|\xi'_{\bq'}(t)\ra = \bVh |\bq'\ra~.
\end{align}

Similarly,  we can still define the ground state of the time-dependent Hamiltonian as the normalized state annihilated by all the $\ah^a(t)$:
\begin{align}
\ah^a(t) |\Psi_0(t) \ra = 0 ~.
\end{align}
This state is no longer time-independent, nor does it solve the Schr\"odinger equation, but it does minimize the energy expectation value, and 
we can write it explicitly in terms of the eigenkets $|\xi'_{\bq'}(t)\ra$ of $\bqh'(t)$:
\begin{align}
|\Psi_0(t) \ra & =  \int d^N\bq'   \Psi_0(\bq') |\xi_{\bq'}(t)\ra~,  &
\Psi_0(\bq') & = \pi^{-N/4} \exp \left[ - \bq'^T \bq'/2\right]~,
\end{align}
where the latter expression is the normalized $N$-dimensional Gaussian.  The fluctuation path integral, \eqref{IPI2}, can now be expressed as
\begin{align}
\cI[P;t_f,t_i] & = \int d^N\bq_f  d^N\bq_i \la \Psi_{0}(t_f) | \bq_f \ra Z(\bq_f,\bq_i;t_f,t_i) \la \bq_i |\Psi_{0}(t_i) \ra~.
\end{align}

The general form of the propagator $Z(\bq_f,\bq_i;t_f,t_i)$ is given above in~(\ref{eq:propagatorfinalform}), but to use it to evaluate $\cI[P;t_f,t_i]$ we need to express the states $|\Psi_{0}(t_{i,f})\rrangle$ in the position basis of the $\bqh$.  This is accomplished by using the unitary transformation $\bVh$:
\begin{align}
\Psi_0(\bq,t) = \la \bq | \Psi_0(t) \ra =  \int d^N \bq'  \la \bq | \xi'_{\bq'}(t)\ra  \Psi_0(\bq') 
= \int d^N \bq'_i  \la \bq | \bVh^\dag| \bq'\ra  \Psi_0(\bq') ~.
\end{align}
The matrix element $\la \bq| \bV^\dag |\bq'\ra$ can be evaluated explicitly in terms of the symplectic transformation relating the two sets of variables~\cite{Arvind:1995ab}.  When the $\bC_3$ block of the Williamson transformation is invertible, the relation is\footnote{We will see shortly that the invertibility assumption will not affect our final result for $\cI[P;t_f,t_i]$.}
\begin{align}
\la \bq | \xi'_{\bq'}(t)\ra = \frac{1}{\sqrt{(2\pi i)^N \det \bC_3}}\exp\left\{ \ff{i}{2} \left( \bq'^T \bC_4^T (\bC_3^{T})^{-1} \bq' - 2\bq^T (\bC_3^{T})^{-1} \bq' + \bq^T (\bC_3^{T})^{-1} \bC_1^T \bq\right)\right\}~.
\end{align}
Carrying out the Gaussian integral, we then obtain
\begin{align}\label{gndstatewavefunctions}
\Psi_0(\bq,t) & = \frac{\exp(-\ff{1}{2} \bq^T \bB \bq)}{\pi^{n/4}\sqrt{\det(i \bC_3 \bA)}}~,
\end{align}
where
\begin{align}
 \bA & := \iden - i \bC_3^{-1} \bC_4~, & \bB &: = (\bC_{3}^{T})^{-1} \bA^{-1} \bC_{3}^{-1} - i\bC_1 \bC_3^{-1}~.
\end{align}
Since $\cbC$ is symplectic, $\bA$ and $\bB$ are both symmetric (in general time-dependent) matrices.

Applying this to our matrix element, and using the wavefunctions \eqref{gndstatewavefunctions} and the propagator \eqref{eq:propagatorfinalform}, we find
\begin{align}
\cI[P;t_f,t_i] & =  \frac{\pi^{-n/2}}{\sqrt{\det(\bC_{3i} \bC_{3f} \bA_i \bA^\dag_f)}\sqrt{ (2\pi i)^{N} \det \bM_3}} \int d^N \bq_i d^N \bq_f 
\exp\left\{i S'(\bq_f,\bq_i;t_f,t_i)\right\}~,
\end{align}
where $\bM_3$ is a block of the classical evolution operator, \eqref{classicalevolution}, and $S'$ is a modification of the action $S(\bq_f,\bq_i;t_f,t_i)$ from~(\ref{eq:evaluatedaction}):
\begin{align}
S'(\bq_f,\bq_i;t_f ,t_i) & = \frac{1}{2} \begin{pmatrix} \bq_f^T & \bq_i^T \end{pmatrix} \cbS'
\begin{pmatrix} \bq_f \\ \bq_i \end{pmatrix}~,&
\end{align}
where the $2N\times 2N$ matrix $\cbS'$ is analogous to the $\cbS$ matrix we met before.  A convenient expression for it is
\begin{align}
\cbS' & =
-\cbJ \begin{pmatrix} \iden & -(\bM_4^T + i\bB_i \bM_3^T) \\ \bM_1 + i \bB_{f}^\dag \bM_3& -\iden \end{pmatrix} \begin{pmatrix} \bM_3^{-1} & 0 \\ 0 & (\bM_3^{T})^{-1}\end{pmatrix}~.
\end{align}
Carrying out this Gaussian integral, we conclude
\begin{align}
\cI[P;t_f,t_i] &= \frac{1}{\sqrt{\cG[P;t_f,t_i]}}~, \qquad \textrm{where} \cr
\cG[P;t_f,t_i] &:= \det\left(-\ff{i}{2} \bC_{3i} \bC_{3f} \bM_3 \bA_i \bA^\dag_f \right) \det \cbS'~.
\end{align}

The convenient form of $\cbS'$, together with $\cbC \in \Sp(2N,\R)$ allow us to reduce $\det\cbS'$ to an $N\times N$ determinant:
\begin{align}
\det \cbS'  = \det( \bM_2 + i \bB^\dag_f \bM_4 + i \bM_1 \bB_i - \bB_{f}^\dag \bM_3 \bB_{i} ) (\det \bM_3)^{-1}~,
\end{align}
so that
\begin{align}
\cG = \det\left\{ -\ff{i}{2} \bA^\dag_{f} \bC^T_{3f} \left(\bM_2 + i \bB^\dag_f \bM_4 + i \bM_1 \bB_i - \bB_{f}^\dag \bM_3 \bB_{i} \right) \bC_{3i} \bA_i\right\}~.
\end{align}
We observe that $\cbC \in \Sp(2N,\R)$ implies 
\begin{align}
\bC_{3i} \bA_i &= \bC_{3i} - i\bC_{4i}~,&
\bA^\dag_{f} \bC_{3f}^T & = \bC_{3f}^T + i \bC_{4f}^T~, \nonumber\\
\bB_i \bC_{3i} \bA_i & = -(\bC_{2i}+i\bC_{1i})~,&
\bA^\dag_f \bC_{3f}^T \bB_{f}^\dag & = -(\bC_{2f}^T -i\bC_{1f}^T)~,
\end{align}
so that, as promised above, all factors of $\bC_3^{-1}$ and $\bM_3^{-1}$ disappear from the final expression.  

The final result can be elegantly written in terms of the complex time-dependent matrices
\begin{align}\label{Ddefs}
  \bD_{12} & = \bC_1 + i \bC_2~,&
  \bD_{34} & = \bC_3 + i \bC_4~,
\end{align}
which are invertible and satisfy 
\begin{align}
\label{eq:childeroland}
\bD_{34}^T \bD_{12} - \bD_{12}^T \bD_{34}  &= 0~, &
\bD_{34}^T \bD_{12}^\ast-\bD_{12}^T \bD_{34}^\ast & = 2i\iden~.
\end{align}
The result is
\begin{align}
\label{eq:finalPItimedependent}
\cG & = \det\left\{ -\ff{i}{2} \left(\bD_{34f}^T \bM_2 \bD_{34i}^\ast - \bD_{12f}^T \bM_4 \bD_{34i}^\ast + \bD_{34f}^T \bM_1 \bD_{12i}^\ast - \bD_{12f}^T \bM_3 \bD_{12i}^\ast\right) \right\}~.
\end{align}
This $N\times N$ determinant is then a complete solution to the regularized path integral over the fluctuations around the soliton in the time-dependent case.  Notice that it depends only on the blocks of the classical evolution operator, \eqref{classicalevolution}, and on the blocks of the Williamson transformation at the initial and final times, $\cbC(t_{i,f})$.

We can give a nontrivial check of the result by considering the $\dot P = 0$ limit.  In this case the symplectic transformation is time-independent, and the components of the evolution operator are determined in terms of the matrix $\cbR$ appearing in~(\ref{eq:transformedsolution}):
\begin{align}
\begin{pmatrix} \bM_1 & \bM_2 \\ \bM_3 & \bM_4 \end{pmatrix} =
\cbC \cbR \cbC^{-1} = \cbC \begin{pmatrix}
\cos \bnu T & -\sin \bnu T \\ \sin \bnu T & \cos\bnu T
\end{pmatrix} \cbC^{-1}~.
\end{align}
Writing these out explicitly in terms of $e^{\pm i \bnu T}$, we find
\begin{align}
2i \bM_2 & =  -\bD_{12}^\ast e^{i\bnu T} \bD_{12}^T + \bD_{12} e^{-i\bnu T}\bD_{12}^\dag ~, \nonumber\\
2i \bM_4 & = -\bD_{34}^\ast e^{i\bnu T} \bD_{12}^T+ \bD_{34} e^{-i\bnu T} \bD_{12}^\dag ~,\nonumber\\
2i \bM_1 & = +\bD_{12}^\ast e^{i\bnu T} \bD_{34}^T - \bD_{12} e^{-i\bnu T} \bD_{34}^\dag ~, \nonumber\\
2i \bM_3 & =  +\bD_{34}^\ast e^{i\bnu T} \bD_{34}^T - \bD_{34} e^{-i\bnu T} \bD_{34}^\dag~.
\end{align}
Finally, plugging these expressions into $\cG$ and using~(\ref{eq:childeroland}), we find $\cG = e^{i (\tr\bnu) T}$, in agreement with~(\ref{eq:constantPresultI}).

\subsection{Computation of the Spectrum for Constant $P$}\label{sec:spectrumcp}

In the previous three subsections we reduced the regulated path integral over the fluctuations to a finite-dimensional determinant.  We were able to obtain fairly general results that did not rely on the particular properties of the quadratic action.  In this subsection, after some preliminary discussion, we will restrict to the case of constant $P$, where we can get an explicit solution to the spectrum of the $\nu_a$ and therefore give a complete solution for the regulated one-loop effective Hamiltonian.\footnote{As discussed under \eqref{opersabove}, we are neglecting terms in $\cbH$ that are exponentially suppressed in $mL$ at large $L$.  Thus the spectrum we obtain here is expected to receive corrections that are analogously suppressed at large $L$.}  This computation serves as a check of our methods and will also be of great utility when these methods are extended to computations of nontrivial matrix elements that contain insertions of the fields.  

The computation is based on applying the Ansatz given in the appendix of~\cite{Christ:1975wt} to our quadratic Hamiltonian, and it hinges on finding the normal modes of the generator of the classical evolution operator, which appears in the equations of motion, \eqref{classicaleoms}.  Thus we are interested in the eigenproblem
\begin{equation}
(\cbJ \cbH) \boeta^a  = - i \nu_a \boeta^a ~,
\end{equation}
for the eigenvectors $\boeta^a$ with normal mode frequencies $\nu_a$.

We begin with some general observations on the eigenvalue problem for the operator 
\begin{equation}\label{AAprimedef}
\cbA' := i \cbJ \cbH~,
\end{equation}
for a positive-definite symmetric $\cbH$.  As the Williamson transformation shows explicitly, this operator is similar to the invertible Hermitian operator
\begin{align}\label{AAdef}
\cbA & := \cbC^{-1} \cbA' \cbC = \begin{pmatrix} 0 & -i\bnu \\ i \bnu & 0\end{pmatrix}~.
\end{align}
Thus, the eigenvalues of $\cbA'$ are real and come in pairs $\pm \nu_a$, $\nu_a > 0$, with complex conjugate eigenvectors $\boeta^{\prime a}$ and $\boeta^{\prime a\ast}$, where $\boeta^{\prime a}$ is the eigenvector corresponding to $+\nu_a$.  These eigenvectors are given by $\boeta^{\prime a} = \cbC \boeta^{a} $ and $\boeta^{\prime a \ast} = \cbC \boeta^{a \ast} $, where $ \boeta^{a}$ is the eigenvector of $\cbA$ with eigenvalue $+\nu_a$.

The monic polynomial $Q(\nu) = \det (\nu - i \cbJ \cbH)$ can be factored as $Q(\nu) = R(\nu) R(-\nu)$, where $R(\nu)$ has positive real roots $\nu_a$.  While $Q(\nu)$ is obtained algebraically once $\cbH$ is known, it is not so simple to determine $R(\nu)$.  So, even though our path integral only depends on the sum of the eigenvalues $\Tr \bnu$,  we cannot evaluate this sum in a simple algebraic fashion from $Q(\nu)$.

Moreover, we saw that in the time-dependent case the form of $\cG(T)$ in~(\ref{eq:finalPItimedependent}) explicitly depends on the details of the symplectic transformation (at initial and final times, in neighborhoods of which $P$ is assumed to be constant), so it would be useful to have expressions for components of $\cbC$ in terms of the solution to this eigenvalue problem.  To obtain such an expression, we note that our original coordinates on the phase space use the real basis vectors $\bpi^n$ and $\bxi^n$, given explicitly by
\begin{equation}
\bpi^n  = \begin{pmatrix} \buppsi^n \\ 0 \end{pmatrix} ~, \qquad \bxi^n = \begin{pmatrix} 0 \\ \buppsi^n \end{pmatrix}~,
\end{equation}
where $\buppsi^n$ are our orthonormal basis vectors: $(\buppsi^n)_m = \delta^{n}_{m}$.  Then any vector $\bz \in \R^{2N}$ can be written as
\begin{align}
\bz & = \sum_{n=1}^N \left( p_n \bpi^n + q_n \bxi^n \right)~.
\end{align}
This is simply a rephrasing of \eqref{accelmodeexp} with the mode sum restricted to $1 \leq n \leq N$.  The phase space basis vectors satisfy
\begin{align}
\cbJ \bpi^n  & = \bxi^n~,&
\cbJ \bxi^n & = - \bpi^n~,
\end{align}
and $ (\bpi^m)^T \bpi^n = (\bxi^m)^T \bxi^n = \delta^{mn}$, while $ (\bpi^m)^T \bxi^n = 0$.  In terms of this real basis, the normalized eigenmodes of $\cbA$, \eqref{AAdef}, are simply
\begin{equation}
\boeta^{a} = \ff{1}{\sqrt{2}}  \left( \bpi^a + i \bxi^a \right)~, \qquad \boeta^{a \ast} = \ff{1}{\sqrt{2}}  \left( \bpi^a - i  \bxi^a \right)~,
\end{equation}
corresponding to eigenvalues $+\nu_a$ and $-\nu_a$ respectively.

The eigenvectors of $\cbA'$, \eqref{AAprimedef}, are then given by the Williamson transformation: $\boeta^{\prime a}  = \cbC \boeta^{a}$ and $\boeta^{\prime a \ast} = \cbC \boeta^{a \ast}$.  Introducing the new basis $\bpi^{\prime a}$ and $\bxi^{\prime a} $ such that
\begin{equation}\label{pixiprime}
\boeta^{\prime a} = \ff{1}{\sqrt{2}} \left( \bpi^{\prime a} + i \bxi^{\prime a} \right)~, \qquad  \boeta^{\prime a \ast} = \ff{1}{\sqrt{2}} \left(  \bpi^{\prime a} - i \bxi^{\prime a} \right)~,
\end{equation}
we have
 \begin{align}\label{eq:componentWilliamson}
&  \bpi^{\prime a}  =  \begin{pmatrix} \bC_1 \buppsi^a  \\[1ex] \bC_3 \buppsi^a  \end{pmatrix} ~, \qquad  \bxi^{\prime a}  =\begin{pmatrix} \bC_2 \buppsi^a  \\[1ex] \bC_4 \buppsi^a \end{pmatrix}~,
 \end{align}
in terms of the block components of the Williamson transformation.  Since the Williamson transformation is symplectic, it follows from \eqref{pixiprime} that
 \begin{align}\label{pixiprimenorm}
 (\bxi'^a)^T \cbJ \bpi'^b  & = \delta^{ab}~,&
(\bxi'^a)^T \cbJ \bxi'^b  &  = 0~,&
( \bpi'^a )^T \cbJ  \bpi'^b & = 0~,
 \end{align}
as well as
\begin{align}
(\bxi'^a)^T \cbH  \bxi'^b &=  ( \bpi'^a)^T \cbH  \bpi'^b   = \nu_a \delta^{ab}~,&
(\bxi'^a)^T \cbH \bpi'^b & = 0~.
\end{align}
Hence the transformed basis diagonalizes the Hamiltonian, as advertised in \eqref{HHWdiag}.

Conversely, if we obtain the complete set of normalized eigenvectors of $\cbA' = i \cbJ \cbH$, we can extract the components of the Williamson transformation.  First we take real and imaginary parts to get $\bpi^{\prime a}$ and $\bxi^{\prime a}$ according to \eqref{pixiprime}, and then from \eqref{eq:componentWilliamson} we infer
\begin{align}\label{Williamsoncomps}
& C_{1}^{na} = (\bpi^n)^T \bpi^{\prime a} ~,  \qquad C_{2}^{na} = (\bpi^n)^T \bxi^{\prime a}~, \qquad  C_{3}^{na} = (\bxi^{n})^T  \bpi^{\prime a} ~,  \qquad C_{4}^{na} = (\bxi^n)^T \bxi^{\prime a} ~.
\end{align}
The normalization condition that must hold on the primed vectors is \eqref{pixiprimenorm}, which is equivalent to the following for the eigenvectors of $\cbA'$:
\begin{equation}\label{eq:etaprimesymplecticnorm}
 (\boeta^{\prime a})^T (i \cbJ) \boeta^{\prime b} = \delta^{ab} ~, \qquad ( \boeta^{\prime a})^T (i \cbJ) \boeta^{\prime b \ast} = 0~.
\end{equation}
In order to compute the combinations $\bD_{12}$ and $\bD_{34}$ that appear in $\cG$, \eqref{eq:finalPItimedependent}, it is more convenient to work with the complex vectors $\boeta^{\prime a}$.  Comparing \eqref{Ddefs} and \eqref{Williamsoncomps} we find
\begin{equation}\label{Dmatrixel}
(D_{12})^{na} = (\bpi^n)^T  \boeta^{\prime a } ~, \qquad (D_{34})^{na} = (\bxi^n)^T \boeta^{\prime a } ~.
\end{equation}

Having reviewed the set up of the eigenproblem for a general Hamiltonian, we now turn to our main interest,  the Hamiltonian in~(\ref{eq:FluctuationHamiltonian}).  The operator $\cbA'$ takes the form
\begin{align}
\cbA' & = i\begin{pmatrix}
-\cbB^T & - \cbK \\ \cbM & \cbB 
\end{pmatrix}~,
\end{align}
so that the eigenvalue problem is
\begin{align}
\begin{pmatrix}
-\cbB^T & - \cbK \\ \cbM & \cbB 
\end{pmatrix} \begin{pmatrix} \boeta'^a_1 \\ \boeta'^a_2 \end{pmatrix} = -i \nu_a \begin{pmatrix}  \boeta'^a_1 \\ \boeta'^a_2 \end{pmatrix} ~.
\end{align}
Using the second row, we solve for $\boeta'^a_1$:
\begin{align}
\label{eq:solveforetaprime1}
 \boeta'^a_1 &  = - \cbM^{-1} (\cbB + i \nu_a \iden) \boeta'^a_2~,
\end{align}
and we plug this result into the first row to obtain
\begin{align}\label{eta2equation}
\bDelta^a \boeta'^a_2  = 0~,
\end{align}
where the operator $\bDelta^a$ is
\begin{align}
\bDelta^a = \cbK - \cbB^T \cbM^{-1} \cbB + i\nu_a (\cbM^{-1} \cbB -\cbB^T \cbM^{-1}) -\nu_a^2 \cbM^{-1}~.
\end{align}

The computation of $(\Delta^a)_{mn}$ is straightforward, remembering that $\KK_{mn} = \langle \uppsi_m | \KK \uppsi_n \rangle$, \etc\,where $\MM$, $\BB$ and $\KK$ are given in \eqref{opersabove}.  We make repeated use of the completeness of the $\{ |\uppsi_n \rangle \}$ in the form $\sum_{n \neq 0} |\uppsi_n \rangle \langle \uppsi_n | = \mathbbm{1} - |\uppsi_0 \rangle \langle \uppsi_0 |$, and there are several remarkable simplifications.  For example, one finds 
\begin{equation}
(\MM^{-1})_{mn} = \delta_{mn} - \frac{ \langle \uppsi_m | \varphibar' \rangle \langle \varphibar' | \uppsi_n \rangle }{\langle \varphibar' | \varphibar' \rangle} ~.
\end{equation}
Notice the lack of any dependence on the static zero-mode $|\uppsi_0 \rangle$.  We also drop boundary terms from integration by parts involving $\varphibar'(\pm L/2)$ for the same reasons discussed under \eqref{opersabove}.  We find that $(\Delta^a)_{mn} = \langle \uppsi_m | \Delta^a |\uppsi_n \rangle$, with
\begin{align}\label{Delta1}
\Delta^a = \Delta^a_{\text{loc}} + \frac{ | 2\abar \varphibar'' - \dot{\varphibar}' +i\nu_a \varphibar' \rangle \langle 2\abar \varphibar'' - \dot{\varphibar}' +i\nu_a \varphibar' |}{\langle \varphibar' |\varphibar' \rangle}~,
\end{align}
where the local part of the operator takes the form
\begin{align}\label{Deltaloc}
\Delta^a_{\text{loc}} = - (1-\abar^2) \p_\rho^2 + V_{0}^{(2)}(\varphibar) + 2i\abar \nu_a \p_\rho -\nu_a^2 ~.
\end{align}

In order to find an $\boeta_{2}^{\prime a} $ that solves \eqref{eta2equation}, it is sufficient to find $|\eta_{2}^{\prime a} \rangle$ satisfying $\Delta^a |\eta_{2}^{\prime a} \rangle = 0$ and $\langle \uppsi_0 | \eta_{2}^{\prime a} \rangle =0$ because such a vector satisfies
\begin{equation}
| \eta_{2}^{\prime a} \rangle = \sum_{n \neq 0} (\eta_{2}^{\prime a})_n |\uppsi_n \rangle~,
\end{equation}
and
\begin{align}
0 =&~ \langle \uppsi_m |\Delta^a | \eta_{2}^{\prime a} \rangle = \sum_{n \neq 0} \langle \uppsi_m | \Delta^a | \uppsi_n \rangle (\eta_{2}^{\prime a})_n = (\Delta^a)_{mn} (\eta_{2}^{\prime a})_n ~,
\end{align}
with $(\eta_{2}^{\prime a})_n$ the components of $\boeta_{2}^{\prime a}$.  Note that $(\eta_{1}^{\prime a})_n$ and $(\eta_{2}^{\prime a})_n$ are precisely the matrix components $(D_{12})^{na}$ and $(D_{34})^{na}$ appearing in \eqref{Dmatrixel}.

To proceed further we now restrict ourselves to the case of constant $P$, where $\varphibar'$ is time-independent, and 
\begin{align}\label{Delta2}
\Delta^a = \Delta^a_{\text{loc}} + \frac{ | 2\abar \varphibar'' +i\nu_a\varphibar'\rangle \langle 2\abar\varphibar'' +i\nu_a \varphibar'|}{\langle \varphibar' |\varphibar'\rangle}~.
\end{align}

Inspired by~\cite{Christ:1975wt}, we will now find the requisite $|\eta_{2}^{\prime a} \rangle$ and $\nu_a$ explicitly.  To motivate the result, it helps to consider the $\abar \to 0$ limit in which there are three simplifications:  (i) $\varphibar = \phi_0$ is the static soliton solution; (ii) $\varphibar' = \frac{1}{\sqrt{M_0}} \uppsi_0$; (iii) 
\begin{align}\label{a0Delta}
\Delta^a = -\p_\rho^2 + V_{0}^{(2)}(\phi_0) -\nu_a^2(1 - |\uppsi_0\rangle \langle\uppsi_0 | )~.
\end{align}
Comparing this to~(\ref{eq:3}) we recognize a familiar problem, and the solution is simple:  $|\eta'^a_2\rangle = |\uppsi_a\rangle$ for $\beta \neq 0$, and $\nu_a = \omega_a$.  Notice also that $\Delta^a$ annihilates $|\uppsi_0 \rangle$ for all $\nu_a$.  Indeed, viewing the operator as a rank-one modification of the local operator, $\Delta_{\rm loc}^a = -\p_\rho^2 + V_{0}^{(2)}(\phi_0) -\nu_a^2$, we find that the vector providing the modification is in the image of the local operator.  Specifically, $\Delta_{\rm loc}^a |\uppsi_0\rangle = - \nu_{a}^2 |\uppsi_0 \rangle$.  Using this and $\langle \uppsi_0 | \Delta_{\rm loc}^a | \uppsi_0 \rangle = - \nu_{a}^2$, we can write \eqref{a0Delta} as
\begin{equation}
\Delta^a = \Delta_{\rm loc}^a - \frac{ \Delta_{\rm loc}^a | \uppsi_0 \rangle \langle \uppsi_0 | \Delta_{\rm loc}^a }{\langle \uppsi_0 | \Delta_{\rm loc}^a | \uppsi_0 \rangle } ~,
\end{equation}
This makes it clear that the general solution is $|\uppsi_a \rangle + c_a |\uppsi_0 \rangle$, but orthogonality to $|\uppsi_0 \rangle$ sets $c_a = 0$.

Remarkably, even in the case $\abar \neq 0$, the ket that appears in $\Delta^a-\Delta^a_{\text{loc}}$ is in the image of $\Delta^a_{\text{loc}}$.  Namely, since $-(1-\abar^2)\varphibar'' + V_{0}^{(2)}(\varphibar) \varphibar' = 0$~, 
\begin{align}
i \nu_a |2\abar \varphibar'' + i \nu_a \varphibar'\rangle = \Delta^a_{\text{loc}} |\varphibar'\rangle~,
\end{align}
and therefore
\begin{align}
\langle\varphibar' | \Delta^a_{\text{loc}} |\varphibar'\rangle = -\nu_a^2 \langle \varphibar'| \varphibar'\rangle~,
\end{align}
where again we drop boundary terms involving $\varphibar'(\pm L/2)$.  Hence, up to terms exponentially small at large $L$,
\begin{equation}
\Delta^a = \Delta_{\rm loc}^a - \frac{ \Delta_{\rm loc}^a | \varphibar' \rangle \langle \varphibar' | \Delta_{\rm loc}^a }{\langle \varphibar' | \Delta_{\rm loc}^a | \varphibar' \rangle } ~,
\end{equation}
making it clear that $\Delta^a$ annihilates $|\varphibar'\rangle$ for any $\nu_a$.

Hence our task is now clear.  If we can find modes $|\psi_a\rangle$ that are annihilated by $\Delta_{\rm loc}^a$, then they are annihilated by $\Delta^a$ as well.  Furthermore, we can then subtract off a multiple of $|\varphibar'\rangle$ and still have a solution: $|\eta_{2}^{\prime a} \rangle = |\psi_a \rangle - c_a |\varphibar' \rangle$ will be annihilated by $\Delta^a$ for any constant $c_a$.  The coefficient $c_a$ is fixed by requiring orthonormality of $|\eta_{2}^{\prime a} \rangle$ with $|\uppsi_0\rangle$.  Hence our solution will take the form
\begin{equation}\label{eta2primesolution}
|n_{2}^{\prime a} \rangle =  |\psi_a \rangle - \frac{ \langle \uppsi_0 | \psi_a \rangle}{\langle \uppsi_0 | \varphibar' \rangle } |\varphibar' \rangle ~,
\end{equation}
where $|\psi_a\rangle$ must satisfy $\Delta_{\rm loc}^a |\psi_{a} \rangle = 0$ with appropriate boundary conditions at $\rho = \pm L/2$.    

Since $\Delta_{\rm loc}^a$, \eqref{Deltaloc}, is a natural generalization of the corresponding operator with $\abar = 0$, the spectrum will vary smoothly with $\abar$ and it makes sense to seek an Ansatz for $|\psi_a \rangle$ that is based on eigenmodes of the static fluctuation operator, \eqref{eq:3}.  We define a boosted and plane-wave dressed function
\begin{align}\label{boostedmodes}
\psi_{a}(\rho) & = N_{a} \tuppsi_{a} \left(\frac{\rho - \rho_0}{\sqrt{1-\abar^2}}\right) \exp\left[ \frac{i \abar \mu_{a}(\rho-\rho_0)}{\sqrt{1-\abar^2}}\right]~,
\end{align}
where $N_{a}$ is a normalization constant, and we remind the reader that the parameter $\rho_0$ is fixed by the constraint on the background solution: $\langle \uppsi_0 | \varphibar - \phi_0 \rangle = 0$.  Here $\tuppsi_{a}(\rho)$ solves \eqref{eq:3} with frequency $\tomega_{a}$. We have introduced the tilde ($\tuppsi$'s and $\tomega$'s) since we do not assume that the $\tuppsi_a$ satisfy the same boundary conditions as the $\uppsi_{n}$ at the edges of the box.  The questions of boundary conditions and normalization will be addressed momentarily.  We then observe that $\Delta^{a}_{\text{loc}} \psi_{a_n} =0$ if we choose $\mu_a = \tomega_{a}$, and set 
\begin{align}
\label{eq:eigenvalues}
\nu_{a} = \sqrt{1-\abar^2}\, \tomega_{a}~.
\end{align}

The boundary conditions on $\psi_{a}$ and the normalization constant $N_a$ must be determined from the orthonormality condition \eqref{eq:etaprimesymplecticnorm}.  From \eqref{eq:solveforetaprime1} and \eqref{eta2primesolution} we find $(\eta_{1}^{\prime a})_m = \langle \uppsi_m | \eta_{1}^a \rangle$, with
\begin{equation}\label{eta1primesol2}
|\eta_{1}^{\prime a} \rangle  =  - (\abar \pd_\rho + i \nu_a) |\psi_a \rangle + \frac{\abar |\varphibar'' \rangle \langle \uppsi_0 | \psi_a \rangle}{\langle \uppsi_0 | \varphibar' \rangle}  = - \abar \pd_{\rho} |\eta_{2}^{\prime a}\rangle  - i \nu_a |\psi_a \rangle~,
\end{equation}
where we used $\langle 2\abar \varphibar'' + i \nu_a \varphibar' |\psi_a \rangle \propto \langle \Delta_{\rm loc}^a \varphibar' | \psi_a \rangle = 0$.  Then on the one hand, 
\begin{align}\label{etaorthcheck}
( \boeta^{\prime a})^T (i \cbJ)  \boeta^{\prime b} =&~ -i \left( \langle \eta_{1}^{\prime a} | \eta_{2}^{\prime b} \rangle - \langle \eta_{2}^{\prime a} | \eta_{1}^{\prime b} \rangle \right) \cr
=&~ \int_{-L/2}^{L/2} d\rho \left\{ (\nu_a + \nu_b) \psi_{a}^\ast \psi_b + i \abar \left( (\pd_\rho \psi_{a}^\ast) \psi_b - \psi_{a}^\ast \pd_\rho \psi_b \right) \right\}~,
\end{align}
while on the other hand
\begin{align}\label{DeltalocID}
0 =&~ \int_{-L/2}^{L/2} d\rho \left\{ \psi_{a}^\ast (\Delta_{\rm loc}^b \psi_b) - (\Delta_{\rm loc}^a \psi_{a})^\ast \psi_b \right\} \cr
=&~ (\nu_a - \nu_b)  \int_{-L/2}^{L/2} d\rho \left\{ (\nu_a + \nu_b) \psi_{a}^\ast \psi_b + i \abar \left( (\pd_\rho \psi_{a}^\ast) \psi_b - \psi_{a}^\ast \pd_\rho \psi_b \right) \right\} + \cr
&~ - \left\{ (1-\abar^2) \left( \psi_{a}^\ast \pd_\rho \psi_b - (\pd_\rho \psi_{a}^\ast) \psi_b\right) - i \abar (\nu_a + \nu_b) (\psi_{a}^\ast \psi_b) \right\} \bigg|_{-L/2}^{L/2} ~.
\end{align}
Here, as usual, we dropped boundary terms in \eqref{etaorthcheck} that are exponentially small in $L$, but we cannot drop the boundary terms in \eqref{DeltalocID}:~for those $\tuppsi_a$ with $a$ corresponding to scattering states, the $\psi_a$ behave asymptotically as plane waves, and these boundary terms are $O(1)$.  

Comparing the two results, we see that if $a \neq b$,
\begin{equation}
(\boeta^{\prime a})^T (i \cbJ) \boeta^{\prime b}  = \frac{1}{(\nu_a - \nu_b)} \left\{ (1-\abar^2) \left( \psi_{a}^\ast \pd_\rho \psi_b - (\pd_\rho \psi_{a}^\ast) \psi_b\right) - i \abar (\nu_a + \nu_b) (\psi_{a}^\ast \psi_b) \right\} \bigg|_{-L/2}^{L/2} ~.
\end{equation}
Hence the $\cbJ$ orthogonality condition for the $\boeta^{\prime a} $ will hold if we choose \eg\,periodic boundary conditions for $\psi_{a}(\rho)$.  Given the plane-wave dressing factor in \eqref{boostedmodes}, this translates into the following boundary condition for the $\tuppsi_a$:
\begin{equation}\label{boostedbcs}
\tuppsi_{a}(\trho_+) e^{i \abar \tomega_{a} \trho_+} = \tuppsi_{a}(\trho_-) e^{i \abar \tomega_{a} \trho_-}~, \qquad \textrm{where} \quad \trho_{\pm} = \frac{ \pm \ff{L}{2} - \rho_0}{\sqrt{1-\abar^2}} ~.
\end{equation}
Note this means that the scattering wavefunctions $\tuppsi_{a}$ will need to be taken as complex, behaving asymptotically as plane waves rather than sines and cosines.  This result generalizes and provides a different perspective on the boundary conditions employed by Jain \cite{Jain:1990dq} in his calculation of the one-loop correction to the relativistic soliton energy for $\phi^4$ theory.  That calculation will be revisited in the next subsection.  Here we see that the boundary conditions arise from demanding that the transformation from old to new phase space coordinates is symplectic.

Since the boundary conditions on the $\tuppsi_a$ are different to those on the $\uppsi_n$, the spectrum of eigenvalues is different: $\{ \tomega_a \} \neq \{\omega_n \}$.  The $\tomega_a$ depend on $\abar$ through the boundary conditions and as $\abar \to 0$ the boundary conditions coincide.  Hence we may write $\tomega_a = \tomega_a(\abar)$, with $\tomega_a(0) = \omega_a$.  An analogous statement holds for the wavefunctions.  The bound state spectra will be practically identical at large $L$, differing by terms of $O(e^{-m L})$.

Returning to \eqref{etaorthcheck}, if $a = b$ then we find
\begin{align}
(\boeta^{\prime a})^T  (i \cbJ)  \boeta^{\prime a} =&~ 2\nu_{a_n} \int d\rho \psi_a(\rho)^\ast \psi_a(\rho) = 2 \nu_{a} \sqrt{1- \abar^2} \, N_{a}^2 = 2 \tomega_{a} (1-\abar^2) N_{a}^2~.
\end{align}
Hence the normalization constant is taken to be
\begin{equation}\label{modenorm}
N_{a} = \frac{1}{\sqrt{2\tomega_{a} (1-\abar^2)}} ~.
\end{equation}

In summary, we have provided a complete solution to the diagonalization problem in the constant $P$ case, up to corrections to the spectrum and eigenvectors that will vanish exponentially fast in $mL$ as the box size $L \to \infty$.  The eigenvalues are \eqref{eq:eigenvalues}, and the eigenvectors are \eqref{eta2primesolution} and \eqref{eta1primesol2} with \eqref{boostedmodes}, \eqref{boostedbcs}, and \eqref{modenorm}.  The relativistic spectrum is a classic result going back \cite{Christ:1975wt}.  However, as far as we are aware, this is the first time the normal modes have been obtained within the constraint formalism of \cite{Gervais:1975pa}.  The two most important results for the final subsection are the spectrum \eqref{eq:eigenvalues} determined through the boundary conditions \eqref{boostedbcs}.

\subsection{The One-loop Correction for Constant $P$}\label{sec:oneloopcp}

The results of the previous two subsections can be summarized by saying that the one-loop correction to the soliton effective Hamiltonian in the case of constant $P$ is
\begin{equation}\label{Heff0result1}
H_{\rm eff}^{(0)}[P] = \frac{1}{2\overbar\gamma} \sum_{a=1}^{N} \tomega_{a} + \int d\rho V_{\delta m^2}(\varphibar) ~,
\end{equation}
where $\overbar\gamma = (1-\abar^2)^{-1/2}$ is the Lorentz factor, and the $\tomega_{a}$ are $N$ non-zero-mode eigenvalues of $-\pd_{\rho}^2 + V_{0}^{(2)}(\phi_0(\rho))$, acting on functions $\tuppsi_{a}$ of $\rho \in (-L/2, L/2)$ that satisfy the boundary conditions \eqref{boostedbcs}.  This result holds for the theory in the box, up to corrections of order $e^{-m L}$.  $V_{\delta m^2}(\varphibar)$ is the one-loop mass counterterm, obtained from renormalization in the perturbative sector, evaluated on the boosted soliton solution $\varphibar(\rho) = \phi_0(\overbar\gamma \rho)$.

If our goal were to study the cut-off theory in the box, then the remaining tasks would be to relate the total number of modes, $N$, to the momentum cutoff implicit in $V_{\delta m^2}$, specify which modes we are including, and to say something more precise about the $O(e^{-m L})$ corrections.  Our goal, however, is not to study the cut-off theory in the box; rather, it is to study the continuum theory on $\mathbbm{R}$.  In this case we need not worry about the $O(e^{-m L})$ corrections, but we have a different problem:~neither the $L \to \infty$ nor $N \to \infty$ limit of \eqref{Heff0result1} exists!  The trained quantum field theorist is not perplexed.  This is to be expected since we are accounting for the ground state energies of infinitely many degrees of freedom.  This overall energy is meaningless in quantum field theory.  In contrast, only the differences between each mode's contribution to the energy in \eqref{Heff0result1} and that mode's contribution to the unobservable vacuum energy are meaningful.  

The latter arise from the vacuum to vacuum transition amplitude, $\llangle \Omega | \mathcal{Z} | \Omega \rrangle$, which the matrix elements  in \eqref{genfunnew} should have been normalized by.  Having suppressed this factor in \eqref{genfunnew}, it was also suppressed in our definition of the soliton effective Hamiltonian, \eqref{Heffdef}.  Now we make it explicit, rewriting \eqref{ssH2} as
\begin{align}\label{Heffproper}
e^{-i \int dt  H_{\rm eff}[P]} :=&~ \frac{1}{\llangle \Omega |\mathcal{Z}|\Omega\rrangle}\int [D\varphi D\varpi D\lambda D\nu ]\Psi_{0}[\varphi(t_f,\rho)]^* \Psi_{0}[\varphi(t_i,\rho)] e^{i \int dt (\langle \varpi |\dot{\varphi}\rangle - H_T) } ~,
\end{align}
with $H_T$ still given by \eqref{ssH2}.  At one loop, the vacuum $|\Omega\rrangle$ is the state annihilated by all of the annihilation operators in the perturbative sector.  Hence a computation analogous to \eqref{eq:constantPresultI} shows that 
\begin{equation}
\llangle \Omega |\mathcal{Z}|\Omega\rrangle = \exp\left\{ - \frac{i}{2} T \sum_{n=0}^{N} \omega_{n}^{(0)} + O(g^2) \right\} ~,
\end{equation}
where the $\omega_{n}^{(0)}$, $0 \leq n \leq N$ are the lowest $N+1$ frequencies of the perturbative modes in the box.  Specifically, the $(\omega_{n}^{(0)})^2$ are the eigenvalues of  $-\pd_{x}^2 + V_{0}^{(2)}(\llangle \phi \rrangle)$ acting on functions satisfying periodic boundary conditions at $x = \pm L/2$.  We consider $N+1$ degrees of freedom in the perturbative sector because that is how many we are considering in the soliton sector: one collective coordinate and $N$ perturbative fluctuations around the soliton.  Our conventions are that modes are uniquely labeled by their index, so for the free modes there is a two-fold degeneracy in the spectrum for $n > 0$: $\omega_{0}^{(0)} < \omega_{1}^{(0)} = \omega_{2}^{(0)} < \omega_{3}^{(0)} = \omega_{4}^{(0)} \cdots$. Hence, instead of \eqref{Heff0result1} what we really have is
\begin{equation}\label{Heff0result2}
H_{\rm eff}^{(0)}[P] = \half \sum_{n=0}^{N} \left( \overbar\gamma^{-1} \tomega_{a_n} - \omega_{n}^{(0)} \right) + \int d\rho V_{\delta m^2}(\varphibar) \;,
\end{equation}
and the notation $\tomega_{a_n}$ indicates there is an identification between the two sets of modes that remains to be determined.  In particular, one of the $\tomega$'s will be the zero-mode (or what becomes the zero-mode as $L \to \infty$), and we have freely extended the sum over the $\tomega$'s to include the would-be zero-mode since its value is $O(e^{-m L})$.

The $N \to \infty$ limit, followed by the $L \to \infty$ limit, of
\eqref{Heff0result2} should now exist. In fact, this computation was carried out for the relativistic kink in $\phi^4$-theory in \cite{Jain:1990dq}.  We revisit the calculation here since some steps are different.  The reason is that \cite{Jain:1990dq} is based on the operator formalism of \cite{Christ:1975wt}, which does not obtain the soliton sector Hamiltonian through a quantum canonical transformation as in \cite{Gervais:1975pa,Tomboulis:1975gf}, and so treats the soliton momentum differently.\footnote{The one-loop correction to the relativistic kink energy, without an account of regularization and renormalization, was also treated in a BRST formalism in \cite{Aldabe:1993np}.}  For brevity we restrict ourselves to $\phi^4$ theory.  A more general analysis will appear elsewhere.

The field theory potential, $V_0(\phi)$, and soliton solution, $\phi_0(\rho)$, are given for $\phi^4$-theory in \eqref{models}, with the replacement $m_0 \to m$, as discussed around equation \eqref{Vfull}.  The potential for the quantum mechanics problem determining the normal modes, $V_{0}^{(2)}(\phi_0(\rho))$, is the $\ell =2$ member of the P\"oschl--Teller family.  For the theory on $\mathbbm{R}$, there are two bound states---one with $\tomega = 0$ and one with $\tomega = \sqrt{\frac{3}{2}} m$.  There is a continuum of scattering states, which can be labeled by $k \in \mathbbm{R}$, with $\tomega(k) = \sqrt{k^2 + 2m^2}$.  Hence for $k \neq 0$ there is the usual two-fold degeneracy in the energy spectrum.  The special scattering state at $k = 0$ is sometimes referred to as a ``half bound state,'' or ``resonance.''  

The P\"oschl--Teller potentials are reflectionless, such that the transmission coefficient is a pure phase, $e^{i \delta(k)}$.  The phase shift for the $\ell = 2$ model has derivative
\begin{equation}\label{deltaprime}
\frac{d\delta}{dk} = -2\sqrt{2} \left( \frac{m}{k^2 + m^2} + \frac{m}{2 k^2 + m^2} \right)~.
\end{equation}
For waves incoming from the left one has $\uppsi_L(\rho \to -\infty) = e^{i k \rho}$ and $\uppsi_L(\rho \to +\infty) = e^{i k \rho + \delta(k)}$, while for waves incoming from the right, $\uppsi_R(\rho \to -\infty) = e^{-i k \rho + \delta(k)}$ and $\uppsi_R(\rho \to +\infty) = e^{-i k \rho}$.  If we consider left and right modes separately, we should restrict ourselves to $k \geq 0$, and then $\delta(k)$ is uniquely determined by continuity once we take the conventional boundary condition $\delta(\infty) = 0$.  It is convenient, however, to define the right-incoming waves as left-incoming waves with $k < 0$.  Then we define $\delta(-k) = -\delta(k)$ and have a single mode for each $k \in \mathbbm{R}$ with the phase shift given by
\begin{equation}
\delta(k) = - 2\pi + 4\pi \Theta(\ff{k}{m}) - 2 \left( \arctan(\ff{k}{\sqrt{2} m}) + \arctan(\ff{\sqrt{2} k}{m}) \right)~,
\end{equation}
where $\Theta(x) = 0$ for $x < 0$ and $\Theta(x)=1$ for $x \geq 0$.  This $\delta(k)$ satisfies $\delta(k \to \pm \infty) = 0$.  The transmission coefficient is smooth through $k=0$ since the discontinuity of $\delta$ is an integer multiple of $2\pi$.  Indeed, the value $\delta(0) = 2\pi$ is predicted by Levinson's theorem.\footnote{Reflectionless potentials in one dimension are special cases of ``exceptional potentials,'' defined by the property of having a resonance at $k=0$---a nontrivial solution in the continuous spectrum.  For such exceptional potentials, Levinson's theorem gives $\delta(0) = \pi n_{\rm b}$, where $n_{\rm b}$ is the number of bound states.  See \cite{Barton:1984py,MR1861473}.}

For the theory in the box we must impose the boundary conditions \eqref{boostedbcs} on these scattering wavefunctions.  For large $m L$ such that the above asymptotics apply, this leads to a quantization condition determining the allowed wavenumbers $k = k_{s}$:
\begin{equation}\label{quantcon}
\overbar\gamma \left( k_{s} + \abar \sqrt{ k_{s}^2 + 2m^2} \right) L + \delta(k_{s}) = 2\pi s ~,
\end{equation}
for $s \in \mathbbm{Z}$.  The form of this condition can be used to identify the correspondence between perturbative modes and soliton sector modes by considering $k_s \gg m$, so that $\delta(k_s) \to 0$, and the effects of the potential can be ignored.  We then find that
\begin{equation}
\overbar\gamma( 1 \pm \abar) k_s = \sqrt{\frac{ 1 \pm \abar}{1 \mp \abar}} \, k_s \approx \frac{2\pi s}{L} ~,
\end{equation}
where the top (bottom) sign is chosen for $k$ positive (negative).  The prefactor is precisely the relativistic Doppler shift due to the fact that we are working in the moving frame of the soliton.  Indeed, the term proportional to $L$ in \eqref{quantcon} is simply the Lorentz transformation of the momentum back to the lab frame.  Hence we see that the modes labeled by momentum $k_s$ should be identified with the perturbative modes labeled by momentum $q_s = \frac{2\pi s}{L}$.

\begin{figure}[t]
\begin{center}
\includegraphics[width=5cm]{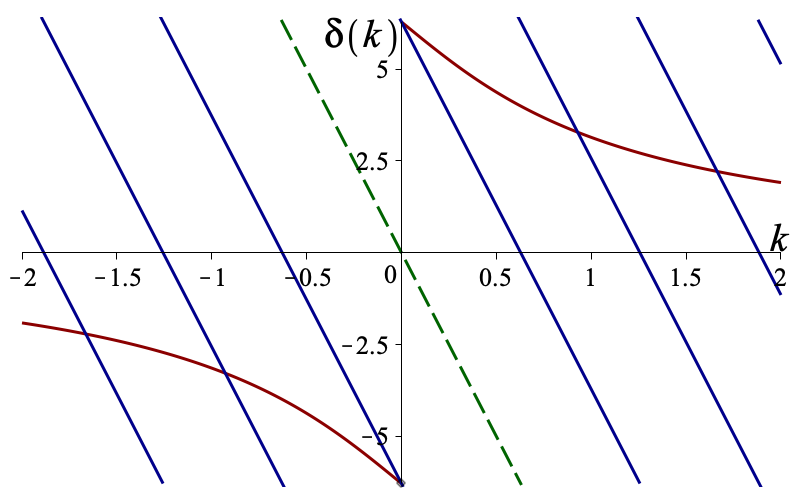}~\includegraphics[width=5cm]{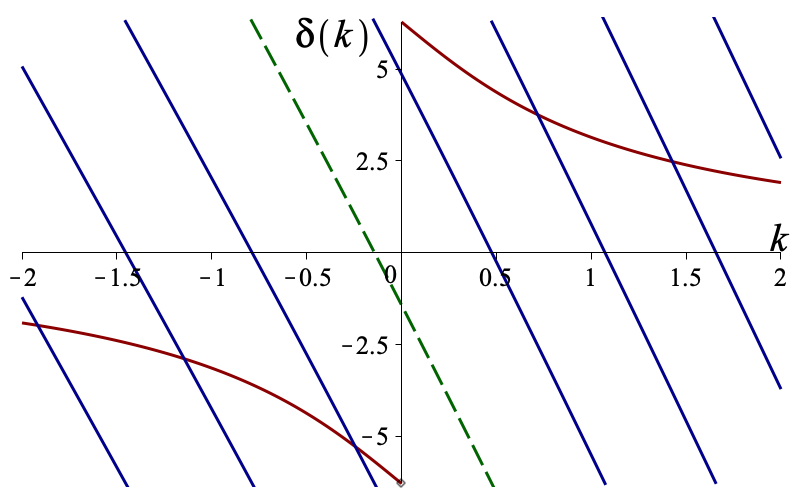}~\includegraphics[width=5cm]{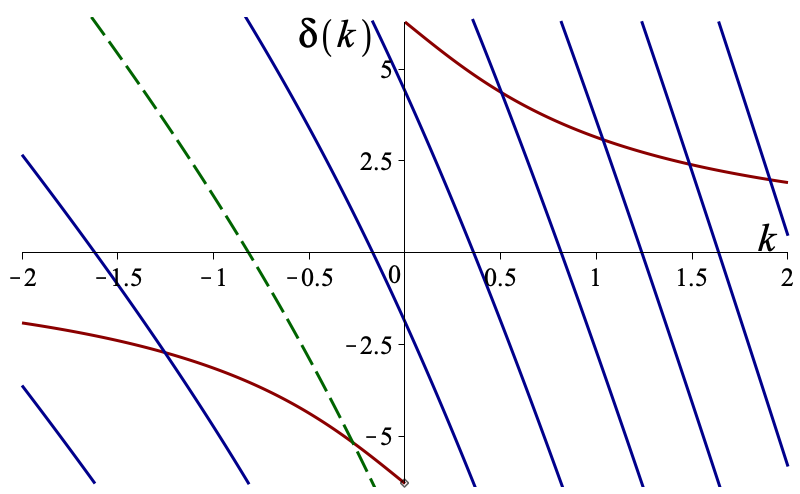}
\end{center}
\caption{$\delta(k)$ (in red) and $y_s(k)$ for various $s$ plotted in units where $m=1$. We set $L=10$ and use $\abar = 0$ (left), $\abar = 1/10$ (center), and $\abar=1/2$ (right).  The dashed green curve is the $s= 0$ curve $y_0(k)$.}
\label{fig1}
\end{figure}

As we decrease $k$, the effects of the potential become important, and we know that two of the modes must be captured by the potential and become the bound states when $L \to \infty$.  We plot $\delta(k)$ and a set of the 
\begin{equation}
y_{s}(k) := 2\pi s - \overbar\gamma \left( k + \abar \sqrt{ k^2 + 2m^2} \right) L ~,
\end{equation}
for several $s$, for three different values of the velocity $\abar$ in Figure \ref{fig1}.  One can see that there are always two neighboring values of $s$ for which there is no solution to $\delta(k) = y_{s}(k)$.  Let us define $s_0$ as the lower of the two integers for which there is no solution and $s_1 = s_0 + 1$ as the higher of the two integers.  When $\abar = 0$ these integers are $s_0 = -1$, and $s_1 = 0$, as pointed out by \cite{Nastase:1998sy}.  As $\abar$ changes, however, the value of $s_1$ (and hence $s_0$) can jump.  This is natural since when $\abar \neq 0$ the mode energies are blue-shifted or red-shifted, and the modes that have the lowest energies will depend on $\abar$.  We can find an expression for $s_1$ by studying the condition $0 \leq y_{s_1}(0) < 2\pi$.  This gives 
\begin{equation}
s_1 = \left\lceil \frac{\abar \overbar\gamma m L}{\sqrt{2} \pi} \right\rceil ~,
\end{equation}
where $\lceil x \rceil$ is the least integer greater than or equal to $x$.  We can conveniently use these two integers to label the bound states: $\tomega_{s_0} = 0$ and $\tomega_{s_1} = \sqrt{\frac{3}{2}} m$.

We can now identify the mode sums appearing in \eqref{Heff0result2}.  We set $N = 2 N_{\Lambda} + 1$, and we assume that for any $\abar$ and $L$, $N_\Lambda \gg |s_1| \sim m L$.  Then we set
\begin{align}
S_{N_\Lambda}^{(0)} :=&~ \sum_{n=0}^{2N_\Lambda + 1} \omega_{n}^{(0)} = \sum_{s= -N_\Lambda}^{N_\Lambda} \sqrt{ ( \ff{2\pi s}{L} )^2 + 2m^2} ~, \label{pertsecsum} \\
S_{N_{\Lambda}} :=&~ \sum_{n=0}^{2N_{\Lambda} + 1} \tomega_{a_n} = \tomega_{s_0} + \tomega_{s_1} + \left( \sum_{s = -N_{\Lambda}}^{s_0 -1} + \sum_{s = s_1 + 1}^{N_{\Lambda}} \right) \sqrt{ k_{s}^2 + 2m^2} ~, \label{solsecsum}
\end{align}
where the $k_s$ are the solutions to \eqref{quantcon}, so that \eqref{Heff0result2} is
\begin{equation}\label{Heff0result3}
H_{\rm eff}^{(0)}[P] = \half \left( \frac{1}{\overbar\gamma} S_{N_\Lambda} - S_{N_{\Lambda}}^{(0)} \right) + \int d\rho V_{\delta m^2}(\varphibar)~.
\end{equation}

To understand the meaning of $N_\Lambda$, consider the sum over the $2N_{\Lambda}+1$ lowest perturbative frequencies.  The largest momentum, $q$, in the sum has $|q| = \frac{2 \pi N_{\Lambda}}{L}$.  Hence we set
\begin{equation}
\Lambda = \frac{2\pi N_\Lambda}{L} 
\end{equation}
and identify this with the UV cut-off from the perturbative calculation of the mass counterterm.  The condition $N_\Lambda \gg mL$ is then simply the condition that $\Lambda \gg m$.  The one-loop mass counterterm for $\phi^4$ theory, evaluated on the soliton background, is
\begin{equation}
\int d\rho V_{\delta m^2}(\varphibar(\rho)) = \frac{\delta m^2}{2} \int_{-\infty}^{\infty} d\rho \left( \llangle \phi \rrangle^2 - \phi_{0}(\overbar\gamma \rho)^2 \right) ~, \qquad \delta m^2 = \frac{3 g^2}{2\pi} \int_{0}^{\Lambda} \frac{dk}{\sqrt{k^2 + 2m^2}} ~.
\end{equation}
Using the explicit soliton solution, one finds
\begin{equation}\label{mctvalue}
\int d\rho V_{\delta m^2}(\varphibar(\rho)) = \frac{3\sqrt{2} m}{2\pi \overbar\gamma} \int_{0}^{\Lambda} \frac{dk}{\sqrt{k^2 + 2m^2}} = \frac{3\sqrt{2} m}{4\pi \overbar\gamma} \, \textrm{Ln}\left( \ff{2\Lambda^2}{m^2}\right) + O \left(\ff{m}{\Lambda^2}\right)~.
\end{equation}

The remaining task is to evaluate the mode sums, at least to sufficiently high order in $m/\Lambda$.  We first use the Euler--Maclaurin formula to convert the sums to integrals $\int ds$, and then change variables to turn them into integrals over the momentum.  For the Euler--Maclaurin formula it turns out to be sufficient to keep the leading boundary terms, 
\begin{equation}
\sum_{s_1}^{s_2} f(s) \to \int_{s_1}^{s_2} ds f(s) + \half (f(s_1) + f(s_2))~, 
\end{equation}
as the higher corrections vanish in the limits $L,\Lambda \to \infty$.  The sum over perturbative sector frequencies is then
\begin{equation}
S_{N_{\Lambda}}^{(0)} = \Lambda + \frac{L}{2\pi} \left( \Lambda^2 + m^2 \, \textrm{Ln}\left( \ff{2\Lambda^2}{m^2}\right) + m^2\right) + O\left( \ff{1}{\Lambda},\ff{1}{L}\right)~.
\end{equation}

For the soliton sector sum we apply Euler--Maclaurin to each sum in the last line of \eqref{solsecsum} separately.  The first step is to determine the values of $k_{s}$ at the four boundary values of $s$.  At large $mL$ the inner boundary values at $s = s_{0}-1, s_1 + 1$ can be found by approximating $\delta(k) = \pm 2\pi + O(k/m)$, for $k \to 0_{\pm}$, which results in
\begin{align}\label{innerboundaries}
k_{s_1 + 1} =&~ \frac{2\pi}{\overbar\gamma L} \left(  \left\lceil \frac{\abar \overbar\gamma m L}{\sqrt{2} \pi} \right\rceil - \frac{\abar \overbar\gamma m L}{\sqrt{2} \pi} \right) + O \left( \ff{1}{m L^2}\right)~,  \cr
k_{s_0 - 1} =&~ \frac{2\pi}{\overbar\gamma L} \left(  \left\lceil \frac{\abar \overbar\gamma m L}{\sqrt{2} \pi} \right\rceil - \frac{\abar \overbar\gamma m L}{\sqrt{2} \pi} -1\right) + O \left( \ff{1}{m L^2}\right) ~.
\end{align}
Meanwhile for $|s| \gg mL$, $|k_s|/m$ will be large, and we can solve \eqref{quantcon} approximately by using the asymptotic value of the phase shift:
\begin{equation}
\delta(k) = \frac{3\sqrt{2} m}{k} + O\left( (\ff{m}{k})^3\right)~.
\end{equation}
Defining $\Lambda_{\pm} := k_{\pm N_\Lambda}$, we find
\begin{equation}\label{outerboundaries}
\Lambda_{\pm} = \frac{ \pm \Lambda}{\overbar\gamma (1\pm \abar)} - \frac{\overbar\gamma \abar m^2}{\Lambda} \mp \frac{3\sqrt{2} m}{L \Lambda} + O \left( \ff{1}{(L\Lambda)^2}\right)~
\end{equation}
where we note  the absence of an $O(\Lambda^0)$ term.  From \eqref{quantcon} we have
\begin{equation}
\frac{ds}{dk} = \frac{\overbar\gamma L}{2\pi} \left(1 + \frac{\abar k}{\sqrt{k^2 + 2m^2}} \right) + \frac{1}{2\pi} \frac{d\delta}{dk}~
\end{equation}
and thus obtain
\begin{align}
S_{N_{\Lambda}} =&~ \cI_1 + \cI_2 + \overbar\gamma \Lambda + \sqrt{2} m + \sqrt{\ff{3}{2}} m+ O\left( \ff{1}{\Lambda},\ff{1}{L}\right)~,
\end{align}
where
\begin{align}
\cI_1 =&~ \frac{\overbar\gamma L}{2\pi} \left( \int_{\Lambda_-}^{k_{s_0 -1}} + \int_{k_{s_1 +1}}^{\Lambda_+} \right) dk \left( \abar k + \sqrt{k^2 + 2m^2} \right)~, \cr
\cI_2 =&~ \frac{1}{2\pi} \left( \int_{\Lambda_-}^{k_{s_0 -1}} + \int_{k_{s_1 +1}}^{\Lambda_+} \right) dk  \sqrt{k^2 + 2m^2} \, \frac{d\delta}{dk} ~.
\end{align}

The integrals can be evaluated, using \eqref{deltaprime} in the case of $\II_2$.  The results can be expanded for large $\Lambda,L$ using \eqref{innerboundaries} and \eqref{outerboundaries}.  The computation of $\mathcal{I}_1$ is delicate.  It is quadratic in $\Lambda$ and proportional to $L$, so all subleading terms we have displayed in \eqref{innerboundaries} and \eqref{outerboundaries} are potentially relevant.  The result is
\begin{equation}
\cI_1 = \frac{\overbar\gamma L}{2\pi} \left\{ \Lambda^2 + m^2 \, \textrm{Ln}\left( \ff{2\Lambda^2}{m^2}\right) +m^2 - \frac{6\sqrt{2} m}{\overbar\gamma L} - \frac{2\pi \sqrt{2} m}{\overbar\gamma L} \right\} + O\left( \ff{1}{\Lambda},\ff{1}{L}\right) ~.
\end{equation}
The evaluation of $\mathcal{I}_2$ is more straightforward.  Since there is no overall factor of $L$, and the inner boundaries are $O(1/L)$, those boundary terms do not contribute and we can integrate directly from $\Lambda_{-}$ to $\Lambda_+$.  Since the integral is logarithmically divergent, we only need the leading behavior of $\Lambda_{\pm}$.  The result is
\begin{equation}
\II_2 = - \frac{3\sqrt{2} m}{2\pi} \, \textrm{Ln} \left( \ff{2 \Lambda^2}{m^2} \right)  - \sqrt{\frac{2}{3}} m +  O\left( \ff{1}{\Lambda},\ff{1}{L}\right) ~.
\end{equation}

Hence, at this order in the large $\Lambda$ and large $L$ expansion, the soliton sector mode sum contains two groups of terms---those proportional to $\overbar\gamma$ and those independent of $\abar$:
\begin{align}
S_{N_\Lambda} =&~ \overbar\gamma \left\{ \Lambda + \frac{L}{2\pi} \left( \Lambda^2 + m^2 \, \textrm{Ln}\left( \ff{2\Lambda^2}{m^2}\right)  + m^2 \right) \right\} - \frac{3\sqrt{2} m}{2\pi } \, \textrm{Ln}\left( \ff{2\Lambda^2}{m^2}\right) + \cr
&~ +  \left(\frac{1}{\sqrt{6}} - \frac{3\sqrt{2}}{\pi} \right)m + O\left( \ff{1}{\Lambda},\ff{1}{L}\right)~. 
\end{align}
The terms proportional to $\overbar\gamma$ are precisely the perturbative mode sum, so these terms completely cancel out of the difference $\overbar\gamma^{-1} S_{N_\Lambda} - S_{N_\Lambda}^{(0)}$, leaving
\begin{equation}\label{modesumdiff}
\half \left( \overbar\gamma^{-1} S_{N_\Lambda} - S_{N_\Lambda}^{(0)} \right) = - \frac{3\sqrt{2} m}{4\pi \overbar\gamma} \, \textrm{Ln}\left( \ff{2\Lambda^2}{m^2}\right) + \left(\frac{1}{2\sqrt{6}} - \frac{3}{\pi\sqrt{2}} \right)\frac{m}{\overbar\gamma} + O\left( \ff{1}{\Lambda},\ff{1}{L}\right)~,
\end{equation}
for the difference that appears in \eqref{Heff0result3}.  The remaining logarithmic divergence in \eqref{modesumdiff} cancels against the mass counterterm.  The limits $\Lambda \to \infty$ and $L \to \infty$ can now be taken, leaving the finite result
\begin{equation}
H_{\rm eff}^{(0)}[P] = \overbar\gamma^{-1} \delta M = \frac{ M_0 \delta M}{\sqrt{ P^2 + M_{0}^2}}~,
\end{equation}
where 
\begin{equation}
\delta M = \left(\frac{1}{2\sqrt{6}} - \frac{3}{\pi\sqrt{2}} \right) m~
\end{equation}
is the one-loop correction to the kink mass first computed in \cite{Dashen:1974cj}.  This result, together with \eqref{Heffm2result}, are consistent with the expansion to $O(g^0)$ of
\begin{equation}
H_{\rm eff}[P] = \sqrt{P^2 + (M_0 + \delta M)^2} ~.
\end{equation}
%

\section{Application and Outlook}\label{sec:FW}

In this paper we carried out the saddle-point approximation to the soliton effective Hamiltonian, \eqref{Heffproper}.  The novelty of our computation is that we made no assumptions about the time derivatives of the soliton momentum.  The tree-level, or $O(g^{-2})$, contribution to the effective Hamiltonian is given in \eqref{HeffTreefinal}.  The one-loop, or $O(g^0)$, contribution is given in \eqref{HeffOneloop} in terms of a quadratic fluctuation path integral, \eqref{IPI}.  Results for that integral were obtained in subsection \ref{sec:PIgen} in terms of classical quantities associated with the quadratic fluctuation Hamiltonian: block components of the classical evolution operator and of the Williamson transformation that diagonalizes the quadratic fluctuation Hamiltonian at initial and final times.

In the absence of insertions or external sources, translation invariance implies that the soliton momentum is conserved, $\dot{P} = 0$.  Lorentz invariance then guarantees that the soliton effective Hamiltonian must reduce to the on-shell relativistic energy, $\sqrt{P^2 + M^2}$, with $M$ the quantum-corrected soliton mass.  We verified that our results, restricted to the case $\dot{P} = 0$, reproduce the tree-level and one-loop contributions to the relativistic energy, where we specialized to the case of $\phi^4$ theory for the one-loop contribution.

In subsection \ref{sec:solFF} we showed how the leading semiclassical behavior of soliton form factors---that is, matrix elements of quantum field theory operators between initial and final soliton states, $| \Psi_{P_{i,f}} \rrangle$---reduces to a matrix element in the collective coordinate quantum mechanics.  The quantum mechanics is governed by the one-loop approximation to the soliton effective Hamiltonian, 
\begin{equation}
H_{\rm eff}[P] \to H_{\rm sc}[P] := H_{\rm eff}^{(-2)}[P] + H_{\rm eff}^{(0)}[P]~.
\end{equation}
If one wishes to obtain results for the semiclassical form factor that are valid for momentum transfers of order the soliton mass, $P_f - P_i \sim O(M)$, then it is necessary to work with the time-dependent $H_{\rm eff}^{(-2)}$ and $H_{\rm eff}^{(0)}$ obtained in this paper.  

A shortcoming of the current work is that fully explicit results for $H_{\rm eff}^{(-2)}[P]$ and $H_{\rm eff}^{(0)}[P]$ hinge on having an explicit solution to the forced soliton equation \eqref{fse}.  This is a second-order quasilinear hyperbolic integro-differential equation.  It generalizes the one obtained in \cite{Gervais:1975pa} for constant $P$ to the case of arbitrary $P(t)$.  We do not currently have explicit solutions beyond those for constant $P$.

Nevertheless, $H_{\rm sc}[P]$ does have one redeeming feature that enables us to carry out the final quantum mechanical path integral in \eqref{spFF}: namely, it is independent of $X$.  This is a consequence of the translation invariance of the underlying theory, and it allows us to obtain an explicit expression for the semiclassical soliton form factor in terms of a generating functional constructed from $H_{\rm sc}[P]$.  We describe this result next.

\subsection{The Generator of Semiclassical Soliton Form Factors}

Equation \eqref{spFF} may be stated in the following way:
\begin{align}\label{PsiOPsi}
\llangle \Psi_{P_f} | \widehat{\mathcal{O}}[\widehat{\pi},\widehat{\phi}] | \Psi_{P_i} \rrangle = \llangle P_{f} | \widehat{\mathcal{O}}_{\rm sc}[\widehat{P},\widehat{X}] | P_i \rrangle ( 1 + O(g))~,
\end{align}
for a Weyl-ordered operator $\widehat{\mathcal{O}}_{\rm sc}[\widehat{P},\widehat{X}]$.  The matrix element on the right is computed by the quantum mechanical path integral with respect to the Hamiltonian $H_{\rm sc}[P]$, and an insertion $\mathcal{O}_{\rm sc}[P,X]$ of phase space variables
\begin{equation}\label{OtoOsc}
\mathcal{O}_{\rm sc}[P,X] := \mathcal{O}[\pibar,\phibar]~.
\end{equation}
The second argument of $\mathcal{O}$ on the right-hand side of \eqref{OtoOsc} is $\phibar(t,x) = \varphibar(t,x - X(t))$, where $\varphibar(t,\rho)$ is a solution to the forced soliton equation, \eqref{fsea}, satisfying the constraint \eqref{fseb}, and is thus a functional of $P$.  Meanwhile, the expression for the first argument, $\pibar(t,x)$, follows from \eqref{newpicx} evaluated on the solution to the forced soliton equation with the aid of \eqref{backgroundsummary}:
\begin{align}
\pibar(t,x) =&~ - \left( \frac{ P + \langle \varpibar | \varphibar' \rangle }{\langle \uppsi_0 | \varphibar\rangle} \right) \uppsi_0(x - X(t)) + \varpibar(t,x-X(t))~,
\end{align}
with $\varpibar(t,\rho) = \dot{\varphibar}(t,\rho) - \abar (\varphibar'(t,\rho) - \langle \uppsi_0 |\varphibar' \rangle \uppsi_0(\rho) )$.

We thus consider the matrix element
\begin{align}\label{QMmatel}
\llangle P_{f} | \fh[\widehat{P},\widehat{X}] | P_i \rrangle :=&~ \frac{1}{2\pi} \int [DX DP] e^{i (P_i X_i - P_f X_f)} \exp \left\{ i \int_{t_i}^{t_f} dt' (P \dot{X} - H_{\rm sc}[P]) \right\} f[P,X]~,
\end{align}
for any Weyl-ordered operator $\fh[\Ph,\Xh]$.  This motivates the definition of the generating functional
\begin{align}\label{scsFF}
\cF_{P_f,P_i}[K,\{F,x\}]  &:= \frac{1}{2\pi} \int [DX DP] e^{i (P_i X_i - P_f X_f)}  \times \nonumber\\
&\qquad\qquad\qquad
\times\exp \left\{ i \int_{t_i}^{t_f} dt' (P \dot{X} - H_{\rm sc}[P] - K P - F (x-X)) \right\} ~, 
\end{align}
in terms of which
\begin{equation}\label{scformfactor1}
\llangle P_{f} | \widehat{f}(\widehat{P},x-\widehat{X}) | P_i \rrangle = \left( f\left[ i \frac{\delta}{\delta K(t)}, i \frac{\delta}{\delta F(t)} \right] \cF_{P_f,P_i}[K,\{F,x\}] \right) \bigg|_{K,F = 0} ~.
\end{equation}
This generating functional will allow us to compute the leading-order-in-$g$ behavior of matrix elements of local operators $\widehat{\mathcal{O}}[\pih,\phih]$ in \eqref{PsiOPsi} defined at a single spacetime point $(t,x)$.  One can generalize to consider insertions at multiple points $(t_n, x_n)$ by introducing additional pairs $\{ x_n, F_n \}$.  We refer to $\FF$ as the \emph{generator of semiclassical soliton form factors}.

The usefulness of this formulation is that the path integral defining the generating functional can be evaluated.  As with the integral over the field theory fluctuations, we implicitly employ midpoint discretization, dividing the interval $(t_i, t_f)$ into $N_t$ subintervals of length $\epsilon$ with $N_t \epsilon = T$.  $X$ integration variables are defined at the grid points: $X_k = X(t_i + k \epsilon)$, so that $X_0 = X_i$ and $X_{N_t} = X_f$.  Momentum variables are defined at the midpoints: $P_k = P(t_i + k (\epsilon - \half))$, for $k = 1,\ldots, N_t$.  Then the path integral measure is
\begin{equation}
[DX DP] = \left(\prod_{k=0}^{N_t} dX_k \right) \left( \prod_{k=1}^{N_t} \frac{dP_k}{2\pi} \right)~.
\end{equation}
The dependence of the integrand of $\FF$ on $X_k$ is a pure plane wave, and so integrating out all $X$ variables produces $N_t +1$ delta functions which can then be used to carry out the $P$ integrations.  Together, the $\delta$ functions enforce Newton's Second Law: $\delta [\dot{P} - F]$.  Since there are $N_t + 1$ $X$-integrations but only $N_t$ $P$-integrations, there will be one delta function left over, which enforces the Impulse-Momentum Theorem: $\delta( P_f - P_i  - \int_{t_i}^{t_f} F(t') dt')$.  The latter can be used to simplify the $x$-dependence of the resulting expression.  Hence the result of the phase space path integration is
\begin{equation}\label{scsFFeval1}
\cF_{P_f,P_i}[K,\{F,x\}] = \delta\left( P_f - P_i  - {\textstyle \int_{t_i}^{t_f}} F(t') dt'\right) e^{-i (P_f - P_i)x} \exp\left\{ -i \int_{t_i}^{t_f} dt' (H_{\rm sc}[\Pbar] + K \Pbar ) \right\} ~,
\end{equation}
where all $F$ dependence is contained in $\Pbar(t)$ obtained as a solution to the Second Law. In the presence of the delta function imposing the Impulse-Momentum Theorem, we can give the following useful expression for $\Pbar(t')$:
\begin{equation}\label{Pbarsmart1}
\Pbar(t') = \half (P_i + P_f) + \half \left( \int_{t_i}^{t'} - \int_{t'}^{t_f} \right) d \tilde{t} F(\tilde{t})~.
\end{equation}
This expression extracts the average value of $F$ from the integral by utilizing $P_f$.  We then note that
\begin{equation}\label{Pbarsmart2}
\frac{\delta \Pbar(t')}{\delta F(t)} = \half (\Theta_{\text{hm}}(t' - t) - \Theta_{\text{hm}}(t-t'))~,
\end{equation}
where the ``half-maximum'' step function satisfies\footnote{To understand the appearance of $\Theta_{\text{hm}}$, note that for any positive $t$ and some test function $f(t)$ we have $\int_0^{t} dt \delta(t) f(t) =\int_{-t}^0 dt \delta(t) f(t)= f(0)/2$.}
\begin{align}
\Theta_{\text{hm}}(z) = \begin{cases}  0~, & z<0 ~,\\ 1/2~, &z = 0~, \\ 1~, &z>0~. \end{cases}
\end{align}
The key simplification that follows from~(\ref{Pbarsmart2}) and is valid under the Impulse-Momentum Theorem delta function is that $\frac{\delta \Pbar(t) } {\delta F(t)} = 0$.

In order to apply \eqref{scsFFeval1} to evaluate the semiclassical form factor, \eqref{scformfactor1}, we need to investigate the functional derivatives of $\FF$ with respect to $K$ and $F$.  The implicit dependence of $\FF$ on $F$ contained in $H_{\rm sc}$ through the solution to the forced soliton equation with $P = \Pbar(t)$ is complicated.  The dependence of $\FF$ on $K$, though, is rather simple and allows for an explicit evaluation of all $K$ derivatives.  Consider the derivative of the term $\int_{t_i}^{t_f} dt' K(t') \Pbar(t')$.  Using \eqref{Pbarsmart1} and \eqref{Pbarsmart2} we obtain
\begin{align}
\frac{\delta}{\delta K(t)} \int_{t_i}^{t_f} dt' K(t') \Pbar(t') =&~ \Pbar(t)~, \cr
\frac{\delta}{\delta F(t)}  \int_{t_i}^{t_f} dt' K(t') \Pbar(t')  =&~  \frac{1}{2} \int_{t_i}^{t_f} dt' K(t') \left( (\Theta_{\text{hm}}(t' - t) - \Theta_{\text{hm}}(t-t')\right)~.
\end{align}
Using either of these we find that the mixed second derivative vanishes
\begin{equation}
\frac{\delta^2}{\delta K(t) F(t)}  \int_{t_i}^{t_f} dt' K(t') \Pbar(t')  = 0~,
\end{equation}
as do all other second and higher order derivatives.  Hence we have that
\begin{equation}
\left( \frac{\delta^{m+n}}{\delta K(t)^m \delta F(t)^n}  \int_{t_i}^{t_f} dt' K(t') \Pbar(t') \right) \bigg|_{K = F = 0} = \left\{ \begin{array}{c l} \ff{1}{2} (P_i + P_f) ~, & m=1 ~\textrm{and}~n=0~, \\ 0~,& \textrm{otherwise}~. \end{array} \right.
\end{equation}

It follows that acting with $K$ derivatives on $\FF$ simply brings down powers of $\half(P_i + P_f)$ such that \eqref{scformfactor1} becomes
\begin{equation}\label{scformfactor2}
\llangle P_{f} | \widehat{f}(\widehat{P},x-\widehat{X}) | P_i \rrangle = \left( f\left[ \ff{1}{2}(P_i + P_f), i \frac{\delta}{\delta F(t)} \right] \cF_{P_f,P_i}[0,\{F,x\}] \right) \bigg|_{F = 0} ~,
\end{equation}
where
\begin{equation}\label{genfunKzero}
\FF_{P_f,P_i}[0,\{F,x\}] =  \delta\left( P_f - P_i  - {\textstyle \int_{t_i}^{t_f}} F(t') dt'\right) e^{-i (P_f - P_i)x} \exp\left\{ -i \int_{t_i}^{t_f} dt' H_{\rm sc}[\Pbar] \right\}~.
\end{equation}
This result is of great practical value since it means that for phase space functions of the form $f = \mathcal{O}[\pibar,\phibar]$, we can use the constant $P = \half (P_i + P_f)$ solution for $\varphibar$ to construct $\pibar,\phibar$.  Thus the differential operator $ f\left[ \ff{1}{2}(P_i + P_f), i \frac{\delta}{\delta F(t)} \right]$ appearing on the right-hand side of \eqref{scformfactor2} will be known explicitly, provided the standard soliton solution is known: $\varphibar(\rho) = \phi_0(\gammabar (\rho-\rho_0))$.  Here $\gammabar$ is the Lorentz factor expressed in terms of the momentum,
\begin{equation}
\gammabar = \sqrt{ 1 + \left( \ff{P_i + P_f}{2M_0}\right)^2} ~.
\end{equation}
Equation \eqref{scformfactor2} is as far as we can go in general without the explicit solution to the forced soliton equation for $P = \Pbar(t)$.  Next we show that \eqref{scformfactor2} reproduces known results in the low momentum transfer limit, with $k = P_f - P_i$ satisfying $|k| \ll M_0$.

\subsection{Semiclassical Soliton Form Factors at Small Momentum Transfer}

Let us consider the $\dot{P}$ expansion of a solution to the forced soliton equation, \eqref{fsea}.  Viewing the forcing term as a perturbation, the leading order solution will be the boosted soliton profile with a boost parameter given in terms of the momentum $\Pbar$.  (Time derivatives of this function will be small and can be grouped with the forcing term as part of the perturbation.)  We can then use our complete knowledge of the diagonalization of the linearized problem around a constant $P$ solution to determine the first perturbative correction proportional to $\dot{P}$.  This first perturbative correction will contribute to the soliton effective Hamiltonian starting at $O(\dot{P}^2)$---either from quadratic terms in the perturbation or from linear terms in the perturbation multiplied by the first order $\dot{P}$ correction to the quadratic fluctuation Hamiltonian, $\cbH$.  Therefore
\begin{equation}\label{HPdotexp}
H_{\rm sc}[\Pbar] = \sqrt{ \Pbar^2 + M^2} + O(F^2)~.
\end{equation}

The $\sqrt{\Pbar^2 + M^2}$ term can also be expanded in $F$ using
\begin{equation}
H_{\rm sc}[F] = H_{\rm sc}[F=0] + \int_{t_i}^{t_f} dt_1 \frac{\delta H_{\rm sc}}{\delta F(t_1)} F(t_1)  + O(F^2)~.
\end{equation}
Since $H$ only depends on $F$ through $\Pbar$, and given the forms of \eqref{HPdotexp} and the derivative \eqref{Pbarsmart2}, it is clear that the $F$ expansion of $H_{\rm sc}[P]$ is also an expansion in $M^{-1}$.  Hence, the leading order term in the $n$-th $F$-derivative of the generating functional \eqref{genfunKzero}, at small momentum transfer, will be given by allowing all $F$-derivatives to act on the delta function factor:
\begin{align}
\left( \frac{\delta^n}{\delta F(t)^n} \FF_{P_f,P_i}[0,{F,x}] \right) \bigg|_{F = 0} =&~  \left( \frac{\delta^n}{\delta F(t)^n} \delta\left( P_f - P_i  - {\textstyle \int_{t_i}^{t_f}} F(t') dt'\right) \right) \bigg|_{F = 0}  \times \cr
&~ \times e^{- i (P_f - P_i)x} e^{-i E_i T} \left(1 + O\left(\ff{P_f - P_i}{M_0} \right)\right) ~,
\end{align}
where the final phase came from $-i \int_{t_i}^{t_f} dt' H_{\rm sc}[F = 0] = -i E_i T$.  Since $E_i = E_f$ to leading order in $(P_f - P_i)/M_0$, we can write this factor as $e^{i E_i t_i - i E_f t_f}$, making it clear that it is the usual normalization associated with asymptotic states, as likewise appears in \eqref{identityFF}.  Hence we have 
\begin{align}\label{scformfactor3}
\llangle P_{f} | \widehat{f}(\widehat{P},x-\widehat{X}) | P_i \rrangle =&~ \left( f\left[ \ff{1}{2}(P_i + P_f), i \frac{\delta}{\delta F(t)} \right] \delta\left( P_f - P_i  - {\textstyle \int_{t_i}^{t_f}} F(t') dt'\right) \right) \bigg|_{F = 0} \times \cr
&~ \times  e^{- i (P_f - P_i)x} e^{-i E_i T} \left(1 + O\left(\ff{P_f - P_i}{M_0} \right)\right) ~.
\end{align}

The form of this result becomes clearer upon considering the Fourier transform with respect to $x$:
\begin{align}
\int dx e^{i k x} \llangle P_{f} | \widehat{f}(\widehat{P},x-\widehat{X}) | P_i \rrangle =&~ \left( f\left[ \ff{1}{2}(P_i + P_f), -i \frac{d}{dk} \right] \delta\left( k  - {\textstyle \int_{t_i}^{t_f}} F(t') dt'\right) \right) \bigg|_{F = 0} \times \cr
&~ \times (2\pi) e^{-i E_i T} \delta(k - (P_f - P_i)) \left(1 + O\left(\ff{k}{M_0} \right)\right) ~,
\end{align}
where we used that $\frac{\delta}{\delta F(t)} \int_{t_i}^{t_f} dt' F(t') = 1$, so that the $\frac{\delta}{\delta F(t)}$ can be replaced with $-\frac{d}{d k}$.  Now we claim that the quantity in the first line of the right-hand side is the Fourier transform of $f \left[ \ff{1}{2}(P_i + P_f), x\right]$.  To see this, consider its inverse transform:
\begin{align}
& \int \frac{dk}{2\pi} e^{-i k x} \left( f\left[ \ff{1}{2}(P_i + P_f), -i \frac{d}{dk} \right] \delta\left( k  - {\textstyle \int_{t_i}^{t_f}} F(t') dt'\right) \right) \bigg|_{F = 0} \cr
& \qquad \qquad =  \int \frac{dk}{2\pi} \delta\left( k  - {\textstyle \int_{t_i}^{t_f}} F(t') dt'\right)  \bigg|_{F = 0} \left( f\left[ \ff{1}{2}(P_i + P_f), i \frac{d}{dk} \right] e^{-i k x} \right) \cr
& \qquad \qquad = \int \frac{dk}{2\pi} \delta(k) f\left[ \ff{1}{2}(P_i + P_f) , x\right] e^{-i k x} \cr
& \qquad \qquad = \frac{1}{2\pi} f\left[ \ff{1}{2}(P_i + P_f) , x\right] ~.
\end{align}
Transforming both sides back then gives the desired relation.  Hence we have arrived at
\begin{align}\label{oldresult}
\int dx e^{i k x} \llangle P_{f} | \widehat{f}(\widehat{P},x-\widehat{X}) | P_i \rrangle =&~ \delta(k - (P_f - P_i)) e^{- i E_i T} \left\{ \int dx e^{i k x} f\left[ \ff{1}{2}(P_i + P_f),x\right] \right\} \times \cr
&~ \times  \left(1 + O\left(\ff{k}{M_0} \right)\right)~.
\end{align}

For the case of the basic field variable itself, $f[P,x] = \phibar(P,x) = \varphibar(\gamma x)$, with Lorentz factor $\gamma = \sqrt{ 1 + (P/M_0)^2}$, the first line on the right side of \eqref{oldresult} matches what was found in \cite{Papageorgakis:2014dma}.  However, the analysis here makes it clear that this result is only the leading order result for the semiclassical form factor in an expansion in $k/M_0$. To obtain an expression for the semiclassical form factor valid when $k \sim O(M_0)$, one must instead use \eqref{scformfactor2} and \eqref{genfunKzero}, which requires solving the forced soliton equation with a time-dependent $P = \Pbar(t)$.  

An interesting output of the result \eqref{oldresult} is that the same Fourier transform prescription gives the leading order behavior of a semiclassical soliton form factor at small momentum transfer for any $f[P,X] = \mathcal{O}[\pibar,\phibar]$.

\subsection{Future Directions}

There are several interesting directions for future work.  We close by listing some of them:     
\begin{itemize}
\item If we are to gain a deeper understanding of soliton form factors at arbitrary momentum transfer, then it seems clear that we must tackle the forced soliton equation head on.  Any progress in this area---existence of solutions, solutions for a special class of $\dot{P}$, numerical exploration---would be helpful.  Although it appears to be a difficult problem, as we stressed in the introduction, the potential implications for our understanding of quantum field theory are deep.  Reference \cite{hughes1977well} on well-posedness of the initial value problem for general second-order quasilinear hyperbolic PDE's appears promising for addressing the question of existence of solutions.
\item A natural first step in a systematic approach towards the forced soliton equation would be to examine perturbative solutions for small but nonzero acceleration.  The results of this paper, including the complete diagonalization of the linearized problem around a constant $P$ solution, leave one well-equipped to address this problem.  Furthermore the results would be new and interesting.  What is the general form of the first higher-derivative corrections to $H_{\rm eff}[P]$ for the class of linear sigma models considered here?  Is there anything that distinguishes the integrable sine-Gordon model from the non-integrable $\phi^4$ model in this regard?  
\item In the case of constant $P$, one could use the explicit diagonalization of the quadratic fluctuation Hamiltonian to set up Feynman rules for perturbative computations around the boosted soliton.  These rules should yield manifestly Lorentz covariant results for $S$-matrix elements in the one-soliton sector, order by order in the coupling expansion.
\item Finally, we would like to extend the analysis conducted here to additional theories admitting solitons, especially gauge theories.  Gauge redundancy presents additional complications for the analog of the canonical transformation, \eqref{cxfmphi} and \eqref{cxfmpi}.  However they are not insurmountable, as shown by the early work of Tomboulis and Woo, \cite{Tomboulis:1975qt}.  This important work needs to be revisited in light of the more geometric approach to gauge theoretic moduli spaces that has been firmly established in the intervening time. 
\end{itemize}

{\bf Acknowledgments:} We would like to thank Shabnam Beheshti, Jim Crawford, Sergei Demidov, Dmitry Levkov, Arick Shao, and Edward Witten for discussions relating to this work. IVM is supported in part by NSF Grant PHY-1914505. The work of CP is supported by a Royal Society University Research Fellowship UF120032 and in part through the STFC Consolidated Grant ST/P000754/1. ABR thanks Queen Mary University of London and James Madison University for hospitality during various stages of this project.



\providecommand{\href}[2]{#2}\begingroup\raggedright\endgroup

\end{document}